\documentclass[twocolumn,preprintnumbers,superscriptaddress,amsmath,amssymb,longbibliography]{revtex4-2}

\usepackage{color}
\usepackage{graphicx}
\usepackage{dcolumn}
\usepackage{float} 
\usepackage{bbm}

\usepackage{setspace}
\usepackage{verbatim}
\usepackage[colorlinks,linkcolor=blue,urlcolor=blue,citecolor=blue,anchorcolor=blue]{hyperref}

\usepackage[dvipsnames]{xcolor}

\def \markColorOne {Plum}
\def \markColorTwo {Red}

\def \markColorOne {Black}
\def \markColorTwo {Black}

\begin{document}
	
	\title{ \color{\markColorOne}  Heisenberg scaling in optical magnetometry with measurement-induced correlations as a quantum resource}
	
	\author{Georg Engelhardt}
	\email{georg-engelhardt-research@outlook.com}
	\affiliation{International Quantum Academy, Shenzhen 518048, China}
	
	\author{Ming Li}
	\affiliation{Shenzhen Institute for Quantum Science and Engineering, Southern University of Science and Technology, Shenzhen 518055, China}
	
	\author{Xingchang Wang}
	\affiliation{Department of Physics, Southern University of Science and Technology, Shenzhen 518055, China}
	
	\author{JunYan Luo}
	\affiliation{Department of Physics, Zhejiang University of Science and Technology, Hangzhou 310023, China}
	
	\author{J. F. Chen}
	\affiliation{Department of Physics, Southern University of Science and Technology, Shenzhen 518055, China}
	\affiliation{International Quantum Academy, Shenzhen 518048, China}
	
	\date{\today}
	
\begin{abstract}
 Theoretical proposals to reach the Heisenberg scaling of the measurement precision typically require carefully engineered interactions or initial entanglement. In studying optical magnetometry,  we show that the continuous collective measurement process itself can generate the necessary many-body quantum correlations to achieve the elusive Heisenberg scaling of the quantum Fisher information  in a dissipative, steady-state system without direct inter-atomic interactions.
	By contrasting a  correlation-neglecting but otherwise consistent semiclassical model, which can violate the quantum Cram\'er-Rao bound (QCRB) by several orders of magnitude, with a collective quantum model, we isolate measurement-induced correlations as the essential mechanism. The violation of the QCRB  serves thereby as a fundamental sanity check for semiclassical spectroscopic theories. This work reveals measurement-induced correlations as a widely unexplored quantum resource for quantum-enhanced sensing, establishes a new paradigm for achieving Heisenberg scaling in open quantum systems, and provides a direct path to test the foundations of quantum mechanics using macroscopic atom ensembles.

\end{abstract}
	
	\maketitle

\begin{figure*}
	\includegraphics[width=\linewidth]{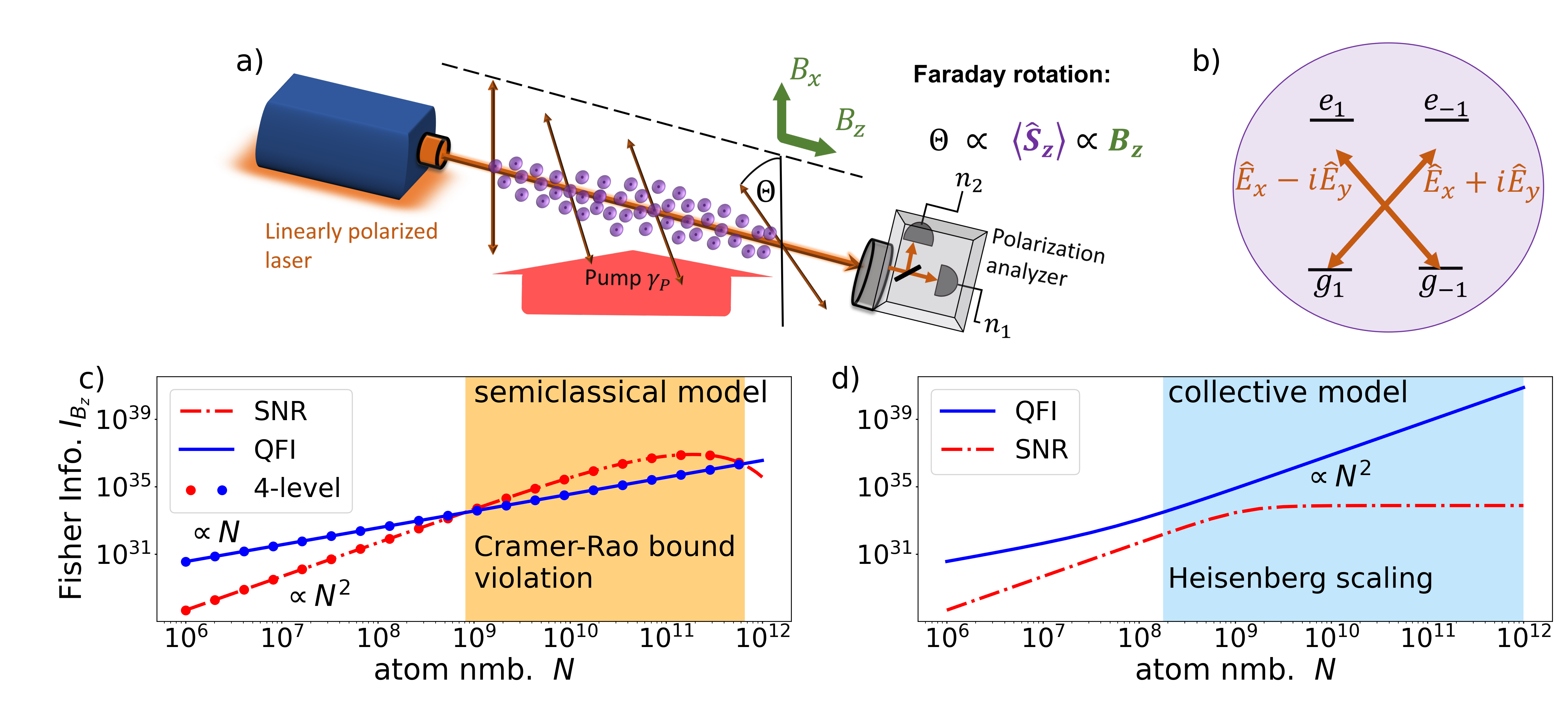}
	\caption{(a) Sketch of the optical magnetometer. (b)  Four-level scheme of the atoms  in the z basis, including transitions induced by the linearly-polarized laser. (c) Signal-to-noise ratio (SNR) and quantum Fisher information (QFI)  in the semiclassical model. Colored lines: Ground-state projection. Dots: semiclassical four-level system. (d) SNR and QFI for the collective model. Parameters are  $d_{\text{A}}  = 4.23 ea_0$ (elementary charge $e$, Bohr radius $a_0$),  $\gamma = 6\,\text{MHz}$, $\gamma_{\text{P}}  = 30\,\text{kHz}$,  $\mu B_{\text{x}}/\hbar = 400 \text{kHz} $ . The probe laser has  wavelength $\lambda = 780\,\text{nm}$, detuning $\epsilon_{\Delta} = \epsilon - \omega_{\text{p}} = 10\,\text{GHz}$, power  $P_{\text{p}} = 10\, \mu\text{W}$, and waist $w_{\text{p}} = 200\,\mu \text{m}$, giving rise to $\Omega = d_{\text{A}} E_{\text{p}}/\hbar =25\,\text{MHz}$. The QFI and SNR are depicted as a function of the atom number $N =\rho_A L \pi w_{\text{p}}^2/4 $ (density $\rho_A$, length $L$), which explicitly determines the optical properties [see, e.g., Eqs.~\eqref{eq:semiclassicalPolarizationRotation}, \eqref{eq:semiclassicalVariance} and \eqref{eq:quantumFisherInfoCollective}].  }
	\label{figOverview}
\end{figure*}
	
\section{Introduction}

	The celebrated quantum Cram\'er-Rao bound (QCRB) sets a fundamental precision limit  in parameter estimation  in terms of the quantum Fisher information (QFI)~\cite{Cramer1946,Rao1945,Helstrom1967,Hayashi2017}. The  Heisenberg limit, referring to the quadratic scaling of the QFI as a function of time or system size, constitutes the ultimate goal in parameter estimation, surpassing  the standard quantum limit, referring to linear scaling, observed in typical measurement protocols~\cite{Giovannetti2004}. The desired Heisenberg scaling has been predicted for several many-body  quantum systems utilizing quantum criticality~\cite{Liu2021,DiCandia2023,Qin2024,Block2024,Sarkar2025},  squeezed or entangled initial states~\cite{Malnou2019,Gessner2020,Backes2021,Mao2023},  precisely coordinated quantum operations~\cite{Kessler2014,Komar2014}, many-body interactions~\cite{Liu2021,DiCandia2023,Qin2024,Block2024,Sarkar2025,Malnou2019,Gessner2020,Backes2021,Mao2023,Kessler2014,Komar2014} and quantum-error correction~\cite{DemkowiczDobrzaifmmodenelsenfiski2014,DemkowiczDobrzaifmmodenelsenfiski2017,Zhou2018}. There is also increasing research attention on  continuous quantum sensing~\cite{Didier2014,Blais2021,Ilias2022,Cabot2024,Qin2024a}. However, these proposals to reach the Heisenberg scaling typically require elaborate experimental control, which is fragile to dissipation.

In this paper, we demonstrate that Heisenberg scaling appears as a robust and inevitable consequence of continuous collective measurements in many-body systems.  As a testbed, we investigate optical magnetometry, an advanced quantum sensing platform~\cite{Bloom1962,Allred2002,Shah2007,Wolfgramm2010,Budker2013,JimenezMartinez2018,Cohen2019,Ma2024,Bao2020,Jin2024,Babaei2025,Hammerer2010,Chen2015,Wang2022,Bloch2022,Afach2021,Brown2025}, which deploys spectroscopic polarization measurement as a continuous collective probe [Fig.~\ref{figOverview}(a) and (b)]. Optical magnetometry has been subject of frequent theoretical investigations~\cite{Thomsen2002,Geremia2003,Moelmer2004,Madsen2004,Zhang2020}, yet, often involving phenomenological assumptions. The measurement backaction of such  continuous  measurements gives rise to  projection noise which sets the standard quantum limit as a generic obstacle for atomic quantum sensors~\cite{Moeller2017,Khalili2018,Jia2023}.

Here, we perform a thorough  quantum information analysis to demonstrate that the inevitable backaction induced correlations can also act as a beneficial quantum resource. We investigate two microscopically derived models: A  semiclassical independent-atom model predicts a drastic violation of the QCRB, heralding its breakdown [Fig.~\ref{figOverview}(c)], as it neglecting quantum correlations generated by the continuous collective measurement. In contrast, a fully collective quantum model—which accounts for measurement-induced correlations due to photon indistinguishably—strictly respects the bound [Fig.~\ref{figOverview}(d)]. Crucially, this restoration is accompanied by Heisenberg scaling {\color{\markColorOne}of the  QFI  } with atom number, emerging not from direct interactions but as an inevitable consequence of the continuous measurement process. As such, the Heisenberg scaling {\color{\markColorOne}of the  QFI  }  appears as an inevitable necessity to maintain the QCRB, whose violation would directly hint to a fundamental problem in the foundations of quantum mechanics. In this context, we  propose  a simple experimental protocol to test the validity of the QCRB.  Thus, our work establishes the correlations induced by the backaction of the continuous collective measurements as a robust quantum resource for achieving the fundamental limit of quantum sensing and testing the foundations of quantum mechanics.

This article is structured as follows. In Sec.~\ref{sec:system}, we explain the system and the Hamiltonian. In Sec.~\ref{sec:quantumCramerRaoBound}, we review the QCRB and specify the measurement observable. In Sec.~\ref{sec:semiclassicalModel}, we investigate the semiclassical model, while in Sec.~\ref{sec:collectiveModel}, we explore the collective model. In Sec.~\ref{sec:discussion}, we discuss our results. Appendices \ref{sec:Hamiltonian}, \ref{sec:fisherInformation}, \ref{sec:app:semiclassicalModel},  and \ref{sec:app:collectiveModel}, provide detailed derivations for the findings in Secs.~\ref{sec:system}, \ref{sec:quantumCramerRaoBound}, \ref{sec:semiclassicalModel}, and \ref{sec:collectiveModel}, respectively.

\section{System}

\label{sec:system}

 The experimental setup of a typical optical magnetometer is sketched in Fig.~\ref{figOverview}(a), where linearly-polarized laser light propagates through a vapor of alkali atoms~\cite{Budker2013}. The Faraday effect due to an external magnetic field in propagation direction induces a rotation of the polarization direction, which is measured by a polarization beamsplitter and two photon detectors.  We microscopically describe the system by the  Hamiltonian~\cite{Hammerer2010}, 
\begin{equation}
\hat  H_{} = \hat  H_{\text{L}} + \sum_m \hat  H_{\text{M}, m} + \sum_m \hat  H_{\text{LM}, m}	 
\label{eq:microscopicHamiltonian},
\end{equation}
where $\hat  H_{\text{L}}$ denotes the Hamiltonian of the free radiation field.  The local Hamiltonian of  atom $m\in\left\lbrace1,\dots N \right\rbrace$, which is modeled   as  a four-level system sketched in Fig.~\ref{figOverview}(b),  is given by
\begin{eqnarray}
\hat  H_{\text{M}, m} &=& \sum_{a=\pm 1}  \epsilon \left| e_{m,a} \right> \left<e_{m,a}\right|   - \sum_{\eta =\text{x,y,z}} \frac{ \mu B_\eta }{2}\hat \sigma_{m,\eta},
\end{eqnarray}
where  $\epsilon$ is  the excited-state energy, and $\left| e_{m,a} \right> $ denote  excited states, with  $a=\pm1$ referring to the quantization in z direction. The ground-state energy is expressed using the common Pauli matrices $\hat \sigma_{m,\eta}$ in the ground-state basis $\left| g_{m,a} \right> $, which is proportional to the product of the external magnetic field $B_{\eta}$ and the magnetic moment $\mu$. For simplicity we assume $B_{\text{y}}=0$. The light-matter interaction in Eq.~\eqref{eq:microscopicHamiltonian} is given by 
\begin{eqnarray}
\hat  H_{\text{LM}, m} 
&=&- d_{\text{A}}  \sum_{a=\pm 1}  \left[   \left| e_{m,-a} \right> \left<g_{m,a}\right| \hat E_{\text{x}  } (\boldsymbol r_m)+\text{h.c.} \right]   \nonumber \\ 
&- &  d_{\text{A}} \sum_{a=\pm 1}  \left[  i \left| e_{m,-a} \right> \left<g_{m,a}\right|\hat E_{\text{y}  }(\boldsymbol r_m)  +\text{h.c.} \right]  \nonumber \\ 
&- & d_{\text{A}}  \sum_{a=\pm 1}  \left[  \left| e_{m,a} \right> \left<g_{m,a}\right|\hat E_{\text{z}  }(\boldsymbol r_m)  +\text{h.c.} \right]  ,
\end{eqnarray}
where  $\hat E_\eta (\boldsymbol r)  $  denotes the $\eta=\text{x},\text{y},\text{z}$ polarization-direction of the  quantized electromagnetic field at position $\boldsymbol r$, and $d_{\text{A}}$ is the transition dipole moment  between the ground and excited-state manifolds. Crucially, this modeling of the atom-light interaction respects   optical selections rules.

{\color{\markColorOne} The coherent probe field of frequency $\omega_{\text{p}}$  is polarized in x direction and propagates in z direction. Details about the microscopic quantization of the probe field and the formal initial condition are given in Appendix~\ref{sec:Hamiltonian}. After propagation through the atom cloud and a polarizing beam splitter [see Fig.~\ref{figOverview}(a)], the probe field intensities in the two polarization directions $\boldsymbol e_{1} = (\boldsymbol e_{\text{y}}  + \boldsymbol e_{\text{y}})/\sqrt{2} $  and $\boldsymbol e_{1} = (\boldsymbol e_{\text{y}} - \boldsymbol e_{\text{y}})/\sqrt{2} $  are measured at the photon detectors $\eta =1$ and $\eta=2$, respectively.  The parameters used in our work mimic existing experimental systems~\cite{Bloch2020,Bloch2022} and are given in the description of Fig.~\ref{figOverview}. }

\section{Quantum Cram\'er-Rao bound} 

\label{sec:quantumCramerRaoBound}

The QCRB sets a fundemental limit on the  estimation precision for a parameter $X$. Invoking the formulation of Ref.~\cite{Gorecki2025},  it reads
\begin{equation}
\mathcal I_X^{(Q)} \geq \mathcal S_{X}^{(A)}, \quad  \text{where} \quad   \mathcal S_{X}^{(A)} \equiv   \frac{\langle  \partial_{X}\hat A   \rangle^2 }{\langle \Delta \hat A^2  \rangle },
\label{eq:FisherInfo-SNR}
\end{equation}
 for any measurement operator $\hat A$, where $\Delta \hat A =  \hat A   -\left<  A \right> $ and $ \mathcal S_{X}^{(A)} $ is the corresponding signal-to-noise ratio (SNR). 
  Thereby, $\mathcal I_X^{(Q)}$ is the QFI, which is defined  as~\cite{Gammelmark2014,Ilias2022}
\begin{equation}
\mathcal I_X^{(Q)}  = 4\int_{0}^{\tau} dt_1\int_{0}^{\tau}dt_2  \left<  \Delta \hat F(t_1) \Delta \hat F(t_2) \right> ,
\label{sec:quantumFisherInfoAtom}
\end{equation}
with measurement time $\tau$, $\hat F (t) = \partial_{X}  \hat H (t)  $, and $ \Delta \hat F(t)   =   \hat F(t) - \langle  \hat F(t) \rangle$, where the time argument refers to the evolution in the Heisenberg picture. For Hamiltonian~\eqref{eq:microscopicHamiltonian} and $X=B_{\text{z}}$, we find $ \hat F =\mu  \sum_m \hat \sigma_{m,\text{z}}/2 \equiv \mu \hat S_{\text{z}} $.
	 
To determine the polarization rotation in optical magnetometry, one often measures the  time-integrated intensities  $\hat n_\eta$   in the two directions $\eta = 1,2$, 
\begin{equation}
\hat n_\eta = \frac{\mathcal A}{\hbar \omega_{\text{p}}}\int_{0}^{\tau}  \hat {\mathcal I}_{\eta} (t) dt,
\label{def:photonNumbers}
\end{equation}
where   $\mathcal A$ is the effective cross section of the probe laser.  The time-integrated intensities are here expressed in units of photon numbers.

 Prior to the  light-matter interaction, the mean time-integrated intensities $\overline n_\eta  = \left< \hat n_\eta \right> $ fulfill  $\overline n_1   =  \overline n_2  $ due to the initial polarization in x direction. We define the measurement operator $\hat A = \hat n_-$ by  
\begin{equation}
\hat n_- \equiv \hat n_{\text{1}}  - \hat n_{\text{2}} 
\label{eq:def:PhotonNumberDifference}
\end{equation}
to detect changes of $B_{\text{z}}$ via the Faraday effect~\cite{Auzinsh2010,Budker2013}. To maximize the measurement sensitivity of the polarization measurement, one often uses a measurement basis rotated by the Faraday induced polarization angle $\overline \theta$, such that the  integrated intensities $\hat n_{\text{rot,1}}$ and $\hat n_{\text{rot,2}}$ fulfill $\langle n_{\text{rot,1}} \rangle = \langle n_{\text{rot,2}} \rangle = \overline n_+/2$. Thus, the mean photon numbers in the original measurement basis are given by
\begin{eqnarray}
\left(
\begin{array}{c}
\overline n_{\text{1}} \\
\overline n_{\text{2}}
\end{array}
\right) 
=
\overline n_{+}	\left(
\begin{array}{c}
\cos^2(\frac{\pi}{4}+ \overline\theta   ) \\
\sin^2(\frac{\pi}{4}+ \overline\theta  )
\end{array}
\right) .
\label{eq:meanPhotonNumbersAngle}
\end{eqnarray}
In the following, we explain how to evaluate the corresponding measurement statistics. This polarization measurement constitutes a continuous collective measurement, probing all atoms simultaneously with a single probe field, such that the photons cannot  distinguish with which atom they have interacted.

\begin{figure}
	\includegraphics[width=\linewidth]{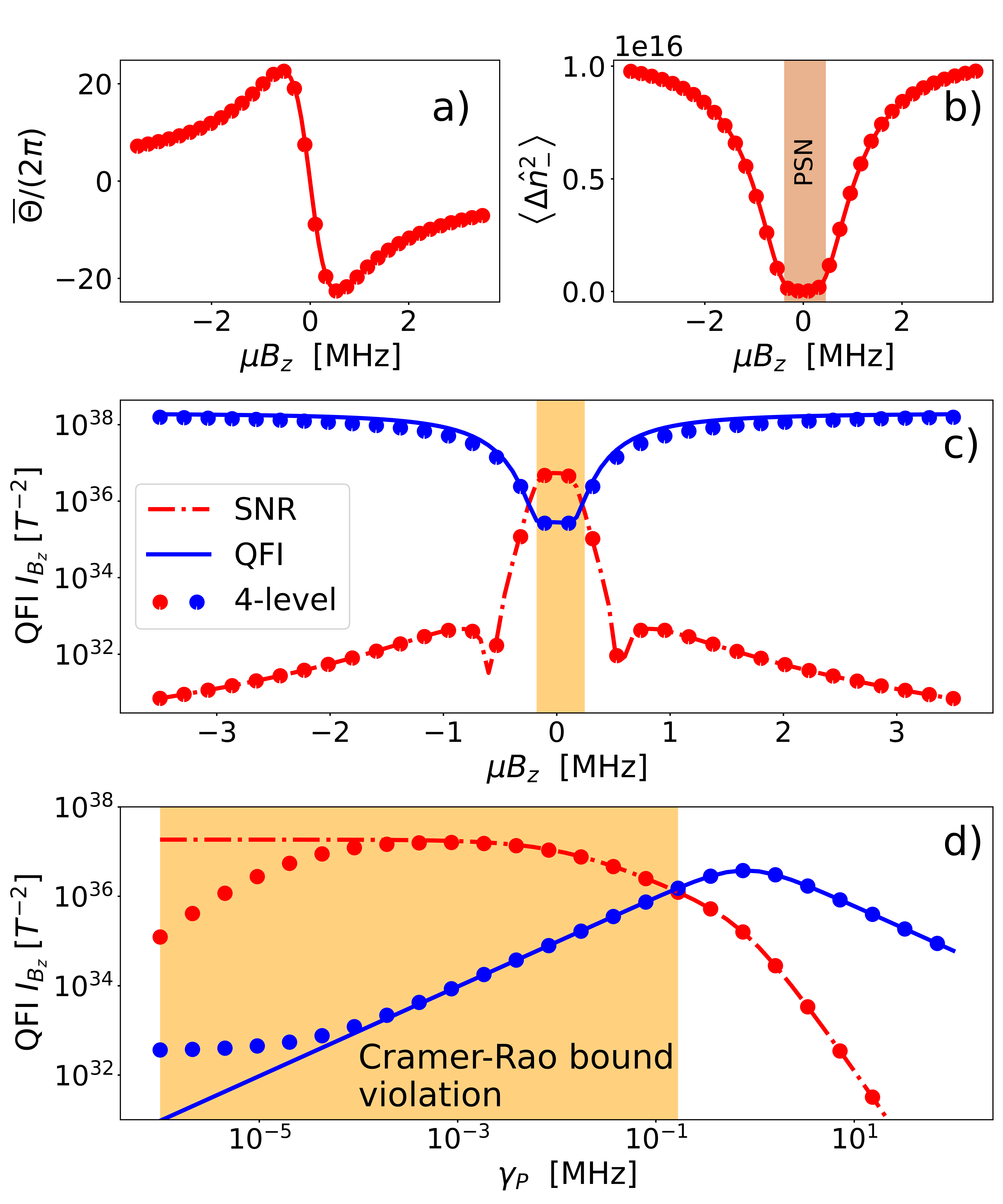}
	\caption{Semiclassical model. (a) Mean polarization rotation in Eq.~\eqref{eq:semiclassicalPolarizationRotation}. (b) Variance of $\hat n_{\text{rot},- }$  in Eq.~\eqref{eq:semiclassicalVariance}. (c) Quantum Fisher information [QFI, blue,  Eq.~\eqref{sec:quantumFisherInfoAtomSC}] and signal-to-noise ratio [SNR, red, Eqs.~\eqref{eq:signal} and~\eqref{eq:semiclassicalVariance}]. (d) QFI and SNR as a function of pumping strength. The colored curves depict  the ground-state manifold master equation in~\eqref{eq:masterEquation}, while the dots show the calculation  for the four-level system. Overall parameters are explained in Fig.~\ref{figOverview}, and $N=8\times 10^{10}$ atoms. }
	\label{figSemiclassicalModel}
\end{figure}

\section{Semiclassical model} 

\label{sec:semiclassicalModel}

{ \color{\markColorOne} 
	 The semiclassical model, which is frequently used in optical magnetometry~\cite{Budker2013,Pustelny2006,Auzinsh2010,Akbar2024,Brown2025}, assumes uncorrelated atoms described by
\begin{equation}
\rho_{\text{M}}  =  \bigotimes_{m=1}^{N} \rho_m ,
\label{eq:productAnsatz}
\end{equation}
where $\rho_m $ denotes the density matrix of atom $m$.  To describe the interaction with the photonic field, we deploy a Markovian quantum-trajectory method to derive a master equation, which guarantees microscopic consistency of quantum information constraints for  each individual atom. We then extend this formalism with the  Maxwell-Bloch theory in a semiclassical fashion to predict the mean and the variance of the measurement statistics of the probe field  after interaction with the full atomic ensemble as shown in Appendix~\ref{sec:app:semiclassicalModel}.  }

For a large detuning of the probe field, the master equation   projected  to the ground-state manifold  reads
\begin{eqnarray}
\frac{d}{dt} \rho
&=&  -i\mu \left[  B_{\text{x}}\hat s_{\text{x}} + B_{\text{z}} \hat s_{\text{z}}  ,\rho \right]  \nonumber \\
&+&  \tilde \gamma_{\text{z}}  D\left[ \hat s_{\text{z}} \right] \rho 
+  	\gamma_{\text{P} } D\left[ \hat s_{\text{z}}- i \hat s_{\text{y}}  \right] \rho ,
\label{eq:masterEquation}
\end{eqnarray}
where $\hat s_\eta =\hat \sigma_\eta/2$ and   $D[ \hat O ]\rho = \hat O \rho \hat O^\dagger -\frac{1}{2}\lbrace \hat O^\dagger\hat O ,\rho  \rbrace$ is the common dissipator. $\tilde \gamma_{z}  =  \frac{\Gamma\Omega^2}{\epsilon_\Delta^2 +  \frac{\Gamma^2}{4} }  $ is the effective dephasing rate resulting from the interaction with the probe laser, i.e., the measurement-induced projection on the measurement basis $\hat s_{z}$. Thereby, $\Omega$ is the Rabi frequency between ground and excited states, $\epsilon_\Delta = \epsilon -\omega_{\text{P} }$ is the detuning, and $\Gamma = \gamma + \gamma_{z}$ is the decay rate from the excited to the grounds states. It is a sum of the spontaneous decay rate in an arbitrary direction $\gamma$, and the decay rate in the direction of the probe-beam direction $\gamma_{\text{z}}$. The latter appears within the quantum-trajectory approach and strictly fulfills $\Omega = 2\sqrt{\gamma_{\text{z}} \overline n_{+}/\tau} $. While it can be neglected in the semiclassical approach as $\gamma_{\text{z}}\ll \gamma $, this parameter plays a crucial role in the collective model due to distinct scaling properties.  We have also phenomenologically introduced a pumping term  $\propto \gamma_{\text{P} }$ polarizing the atom in x direction.

For  large  $\epsilon_{\Delta}$, we  evaluate the measurement statistics using the Maxwell-Bloch formalism as explained in Appendix~\ref{sec:app:semiclassicalModel}. Assuming $\Gamma\Omega^2/\epsilon_{\Delta}^2\ll \gamma_{\text{P}}$ and long measurement times $\tau$,  we obtain the total photon number and  polarization rotation 
\begin{eqnarray}
\overline n_{+} &=& e^{-  \breve I_+  N  } \overline n_{+}^{(i)} ,\nonumber \\
\overline \theta  &=&   \frac{1}{2} \breve I_- N =  N \frac{\mathcal C_{A} \tau}{2\overline n_{+} }   \langle \hat s_{\text{z}}\rangle_{\text{ss}} \nonumber \\
&=&  N \frac{\mathcal C_{A}  \tau}{2\overline n_{+} } \frac{2    h_{\text{x} }  h_{\text{z}}   }{  2 h_{\text{z}}^2    + 4  h_\text{x}^{2} +\gamma_{\text{P}}^2  } ,
\label{eq:semiclassicalPolarizationRotation}
\end{eqnarray}
where $N$ is the number of atoms involved in the light-matter interaction and  $ \overline n_{+}^{(i)}$ is the initial total photon number. The expectation value $\langle \bullet\rangle_{\text{ss}} $ refers to the stationary state.  Thereby,
\begin{equation}
\breve I_\pm  = \frac{1}{\overline n_{+}}\int_{0}^{\tau} \left<  \hat j_{\pm}(t) \right>dt
\label{eq:temporalAverage}
\end{equation}
with $\hat j_{\pm} =\hat j_{\text{1}}  \pm \hat j_{\text{2}} $. Within  the effective ground state model Eq.~\eqref{eq:masterEquation}, the photon flux operators are given by
\begin{eqnarray}
\hat {j}_{\text{1}} &=&  \frac{2\epsilon_\Delta \Omega^2}{ 4\epsilon_\Delta^2 + \Gamma^2   } \hat s_{\text{z}} - \frac{\Gamma \Omega^2}{ 4\epsilon_\Delta^2 + \Gamma^2   },\nonumber \\
\hat { j}_{\text{2}} &=& \frac{-2\epsilon_\Delta \Omega^2}{ 4\epsilon_\Delta^2 + \Gamma^2   } \hat s_{\text{z}} - \frac{\Gamma \Omega^2}{ 4\epsilon_\Delta^2 + \Gamma^2   }.
\label{eq:photonFluxOperatorsSC}
\end{eqnarray}
The last line in Eq.~\eqref{eq:semiclassicalPolarizationRotation} gives an explicit expression for the polarization rotation, where $h_\eta =\mu B_{\eta}$ and $\mathcal C_{A} =\Omega^2/\epsilon_\Delta $, as shown in Appendix~\ref{sec:analyticalCalculations}.  
{\color{\markColorTwo} Here and in the following, we assume asymptotic measurement times which significantly exceed the time scales of the master equation in Eq.~\eqref{eq:masterEquation} [and similarly, the collective master equation in Eq.~\eqref{eq:collectiveMasterEquation} below], namely, $\tau \gg 1/\mu B_{\text{x}} , 1/\gamma_{\text{P}}, 1/ \tilde \gamma_{\text{z}}$.  }

In the large detuning regime, the  response of Eq.~\eqref{eq:def:PhotonNumberDifference} to the magnetic field is given by
\begin{eqnarray}
\left<\partial_{B_{\text{z}}} \hat n_{\text{rot},- }  \right>&=&  \frac{1}{2} \overline n_{+} N \frac{d \breve  I_-}{dB_{\text{z} }}  \label{eq:signal}\\
&=&  \frac{\tau N\mathcal C_{A}     h_{\text{x}} \mu   }{  2  h_{\text{z}}^2    + 4  h_{\text{x}}^{2} +\gamma_{\text{P}}^2  }  
+  \frac{\tau N\mathcal C_{A} 4    h_{\text{x}} h_{\text{z}}^2\mu   }{ \left(  2  h_{\text{z}}^2    + 4  h_{\text{x}}^{2} +\gamma_{\text{P}}^2 \right)^2 },
\nonumber
\end{eqnarray}
and the corresponding noise is
\begin{eqnarray}
\left< \Delta \hat n_{\text{rot},- }^2  \right>  &=&  \overline n_{+} + N \mathcal D \label{eq:semiclassicalVariance} \\ 
 &=& \overline n_{+}  + \tau N \mathcal C_{A}^2   \frac{2 h_{\text{z}}^2   + \gamma_{\text{P}}^2     }{  2 \gamma_{\text{P}} h_{\text{z}}^2    + 4 \gamma_{\text{P}} h_{\text{x}}^{2} + \gamma_{\text{P}}^3 } \nonumber \\ 
&-&2\tau N \mathcal C_{A}^2\frac{ 64 \left(  h_{\text{z}}^2   +  h_{\text{x}}^{2} + \frac{5}{4} \gamma_{\text{P}}^2   \right) \left( \frac{1}{2} \gamma_{\text{P}}  h_{\text{x}}   h_{\text{z}}   \right)^2  }{\left(  2 \gamma_{\text{P}} h_{\text{z}}^2    + 4 \gamma_{\text{P}} h_{\text{x}}^{2} + \gamma_{\text{P}}^3 \right)^3   } ,  
\nonumber
\end{eqnarray}
where $\overline n_{+}$  represents the photon-shot noise (which is equal to the total photon number) and
\begin{equation}
\mathcal D = \int_{0}^{\tau} dt_1\int_{0}^{\tau}dt_2  \left<  \Delta \hat j_{-}(t_1) \Delta \hat j_{-}(t_2) \right> 
\end{equation}
with $ \Delta \hat j_{-}(t)   =   \hat j_{-}(t) - \langle  \hat j_{-}(t) \rangle$ can be interpreted as a  diffusion parameter. The latter can be efficiently evaluated using a full-counting statistics approach as shown in Appendix~\ref{sec:analyticalCalculations} such that we obtain the explicit result in Eq.~\eqref{eq:semiclassicalVariance}.

 Neglecting correlations between distinct atoms in Eq.~\eqref{sec:quantumFisherInfoAtom} according to the semiclassical approximation, we find 
\begin{eqnarray}
\mathcal I_{B_{\text{z}}  }^{(Q)}  &=& 4N \int_{0}^{\tau} dt_1\int_{0}^{\tau}dt_2  \left<  \Delta \hat s_z (t_1) \Delta \hat s_z (t_2) \right> \nonumber \\
 &=& 4\mu^2 	\left( \left< \Delta \hat n_{\text{rot},- }^2  \right> -\overline n_{+}\right)/\mathcal C_{A}^2.
\label{sec:quantumFisherInfoAtomSC}
\end{eqnarray}
This is directly related to the measurement noise in Eq.~\eqref{eq:semiclassicalVariance}, as the Fisher information generator and the polarization operators are determined by $\hat s_{z}$. {\color{\markColorOne}  Crucially, the calculation of the signal in Eq.~\eqref{eq:signal}, the noise in Eq.~\eqref{eq:semiclassicalVariance}, and the QFI in Eq.~\eqref{sec:quantumFisherInfoAtomSC} have been carried out under the separability assumption in Eq.~\eqref{eq:productAnsatz}, which allows a comparison of these quantities on an equal footing. }

Figure~\ref{figSemiclassicalModel} depicts these quantities as a function of  $\mu B_{\text{z}}$ together with the non-adiabatic evaluation using the full four-level system in Eq.~\eqref{eq:microscopicHamiltonian} (while still invoking the semiclassial approach, see Appendix~\ref{sec:app:semiclassicalModel}). In Fig.~\ref{figSemiclassicalModel}(a), the polarization rotation $\overline \theta$ abides to the  well-known dispersion relation, reflecting the relation between the refractive index and the Faraday effect. In Fig.~\ref{figSemiclassicalModel} (b),  the variance $\langle \Delta \hat n_{\text{rot},- }^2  \rangle $  exhibits a flat minimum for $B_{\text{z}}\approx 0$, where the minimal value is close to the photon-shot noise (PSN). For large $\left| B_{\text{z}}\right|$, the variance increases drastically and is  dominated by the second and third terms in Eq.~\eqref{eq:semiclassicalVariance}, which scale  with $1/\gamma_{\text{P} }$ in the weak pumping regime.

To analyze the information content in this setup, we depict the QFI $\mathcal I_{B_{\text{z}   }  }^{(Q)}$ and the SNR  $\mathcal S_{B_{\text{z}   }}^{(n_{-} )}$  in Fig.~\ref{figSemiclassicalModel}(c). For large $\left| B_{\text{z}   }\right|\gg0$, we find that $\mathcal I_{B_{\text{z}}}^{(Q)}$  exceeds $\mathcal S_{B_{\text{z}   }}^{(n_{-} )}$ by several orders of magnitude. This is a result of the time averaging in $\hat n_{\text{rot},-}$ leading to a loss of information~\cite{Radaelli2023,Engelhardt2025}.

Surprisingly, for  $\left| B_{\text{z}}\right|\approx 0$, we observe $\mathcal I_{B_{\text{z}}}^{(Q)} < \mathcal S_{B_{\text{z}   }}^{(n_{-} )} $, meaning that the QCRB is violated. To further investigate this, we depict both quantities in Fig.~\ref{figOverview}(c) as a function of the atom number. The QFI increases linearly with atom number. For small atom numbers, the SNR increases quadratically with the atom number. In this regime, the noise is dominated by the photon-shot noise, which is  independent of the atom number, while the signal in Eq.~\eqref{eq:signal} scales linearly with $N$. Above $N\approx10^{8}$ we observe that  $\mathcal I_{B_{\text{z}} }^{(Q)} < \mathcal S_{B_{\text{z}   }}^{(n_{-} )} $, i.e, the QCRB is violated almost two orders of magnitude. For even larger $N$, $ \mathcal S_{B_{\text{z}   }}^{(n_{-} )}$ becomes flatter and eventually decreases because probe laser absorption becomes significant.

For $B_{\text{z} } =0$ and $N\rightarrow \infty$ we find $\frac{\mathcal S_{B_{\text{z} } }^{(n_{-} )} }{	\mathcal I_{B_\text{z}}^{(Q)}} \rightarrow \frac{\mu B_{\text{x}}}{\gamma_{\text{P}}}$,
 implying that the QCRB  is possibly violated for $\mu B_{\text{x}} > \gamma_{\text{P}}$ which is supported by  Fig.~\ref{figSemiclassicalModel}(d): For the two-level system,  the QCRB is violated by several orders of magnitude for small $\gamma_{\text{P}}$, which is confirmed by the four-level system calculation. Only in the very weak pump regime for which $\gamma_{\text{P} }< \Gamma\Omega^2/\epsilon_{\Delta}^2$, we observe significant deviations between the two- and four-level-system calculations, as there the dissipative dynamics is dominated by  excited-state processes.

The analysis thus demonstrates that a semiclassical QCRB violation is possible for large atom numbers and   weak dissipation. This suggests an emergent collective quantum effect that is not captured by the semiclassial independent-atom model. These collective dynamics eventually give rise to additional contributions to the QFI given by correlations of distinct atoms  in Eq.~\eqref{sec:quantumFisherInfoAtom}.

\begin{figure}
	\includegraphics[width=\linewidth]{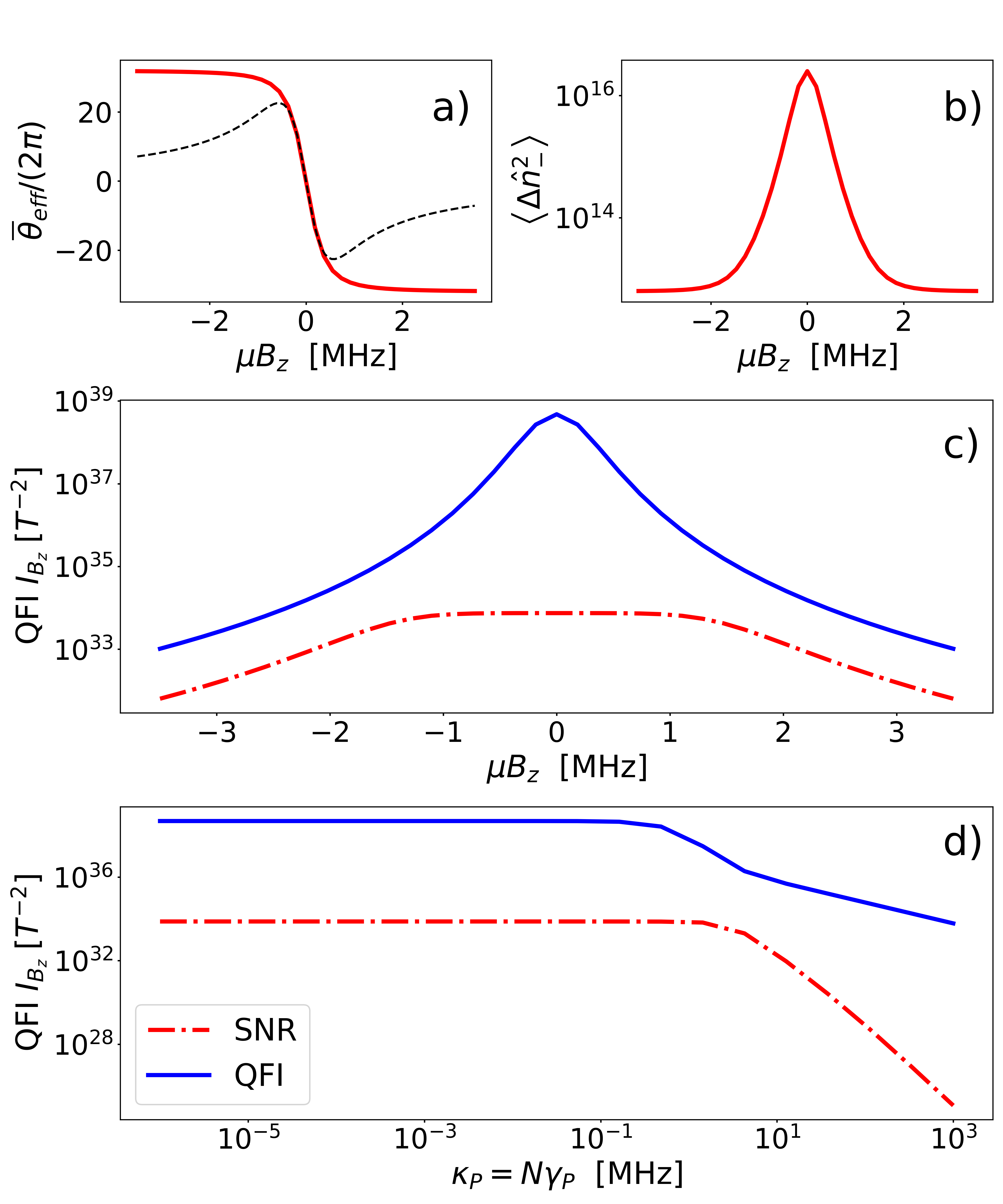}
	\caption{Collective model.  (a) Effective rotation angle (red, solid) compared with the semiclassical rotation angle (black, dashed). (b) Variance of $\hat n_{- }$  in Eq.~\eqref{eq:collectiveNoise}. (c) Quantum Fisher information [QFI, blue] and signal-to-noise ratio [SNR, red]. (d) QFI and SNR as a function of the renormalized  pump strength $\kappa_{\text{P}}$. Overall parameters are the same as in Fig.~\ref{figOverview}, with  $N=8\times 10^{10}$ and  $\kappa_{\text{P} } \equiv \tilde N \gamma_{\text{P} }  = 30\text{kHz}$  }.
	\label{figAnalysisCollective}
\end{figure}

\section{Collective model}

\label{sec:collectiveModel}

 In terms of the collective spin operators $\hat S_\alpha = \sum_m \hat s_{m,\alpha}$, the master equation  in the collective ground-state manifold reads
\begin{eqnarray}
\frac{d}{dt} \rho   &=&  -i\mu \left[   B_{\text{x}} \hat S_{\text{x}} +    B_{\text{z}} \hat S_{\text{z}}  ,\rho \right]  \nonumber \\
&+&  \tilde \gamma_{\text{z}}  D[ \hat S_z ] \rho
+    \gamma_{\text{P} }  D [  \hat S_{\text{z}} -i   \hat S_{\text{y}}   ] \rho 
\label{eq:collectiveMasterEquation}
\end{eqnarray}
with $\tilde \gamma_{\text{z}}$ defined below Eq.~\eqref{eq:masterEquation}.  Here, we  assume scaled $\gamma =\tilde  \gamma/N\approx 0$ and $\gamma_{\text{P}} =\kappa_{\text{P}}/N$ to  obtain a well-defined semiclassical limit $N\rightarrow \infty$ for a meaningful comparison with the semiclassical model. { \color{\markColorTwo}  A more detailed justification of these scaling assumptions can be found at the end of this section.} Crucially, to ensure the validity of the QCRB, $\Omega =2 \sqrt{\gamma_{\text{z}} \overline n_{+}/\tau} $, such that the second term in Eq.~\eqref{eq:collectiveMasterEquation} encodes the collective measurement backaction induced by the magnetic-polarization measurement. Note that Eq.~\eqref{eq:collectiveMasterEquation} generalizes the description  using collective Stokes operators and Bogoliubov transformation, which is often deployed to investigate quantum precision limits of atomic interferometers~\cite{Moeller2017,Khalili2018,Jia2023,Novikov2025,Moelmer2004,Zhang2020,Jia2023,Novikov2025}.

The measured photon numbers in a fixed measurement directions $\eta =\text{1,2}$ can be calculated via 
\begin{eqnarray}
\left<  \hat n_{\eta } \right>  &=&\frac{\overline n_+ }{2} + \int_{0}^{\tau}dt  \left<  \hat {\mathcal J}_{\eta}(t)  \right>    ,
\label{eq:mean:collective}
\end{eqnarray}
with the photon-flux operators $ \hat {\mathcal J}_{\text{1},2} =\pm \hat S_{\text{z}} \Omega^2/2\epsilon_{\Delta} $.  Figure~\ref{figAnalysisCollective}(a) depicts the effective rotation defined as $\overline \theta_{\text{eff}} = (\overline n_{\text{1}} - \overline n_{\text{2}}  ) /2(\overline n_{\text{1}} + \overline n_{\text{2}}  ) = \frac{\mathcal C_{A} \tau}{2\overline n_{+} }   \langle \hat S_z\rangle_{\text{ss}}  $ in the stationary state, which is  formally equivalent to the semiclassical rotation $\overline \theta$ in Eq.~\eqref{eq:semiclassicalPolarizationRotation}. Both $\overline \theta$ and $\overline \theta_{\text{eff}} $ agree for $B_{\text{z}}\approx 0$, justifying a comparison of both models.   Moreover,
\begin{eqnarray}
\left< \Delta \hat n_{- }^2  \right>  &=& \overline n_{+}+\int_{0}^{\tau} dt_1\int_{0}^{\tau}dt_2  \left<  \Delta \hat {\mathcal J}_{-}(t_1) \Delta \hat {\mathcal J}_{-} (t_2) \right> , \nonumber\\  
\label{eq:collectiveNoise}
\end{eqnarray}
with $ \Delta \hat {\mathcal J}_{-}(t)  =    \hat {\mathcal J}_{-}(t) - \langle   \hat {\mathcal J}_{-}(t)\rangle$ and $\hat {\mathcal J}_{-}=\hat {\mathcal J}_{\text{1}} -\hat {\mathcal J}_{\text{2}}$, which can be efficiently evaluated using a non-unitary mean field theory valid in  the thermodynamic limit $N \rightarrow \infty$ as shown in Appendix~\ref{sec:nonUnitaryMeanFieldTheory}~\cite{Li2025}.

The measurement noise in Fig.~\ref{figAnalysisCollective}(b) disagrees from the semiclassical measurement noise in Fig.~\ref{figSemiclassicalModel}(b), as we consider here $\hat n_{-}$ instead of $\hat n_{\text{rot},-}$. Crucially, the overall noise for   $\left| B_{\text{z}} \right|\approx 0$ is three orders of magnitudes larger than for the semiclassical description due to the collective character of the dissipation in Eq.~\eqref{eq:collectiveMasterEquation} generating correlations between the atoms. The SNR $\mathcal S_{B_{\text{z}   }}^{(n_{-} )} $ in Fig.~\ref{figAnalysisCollective}(c) is significantly smaller close to $B_{\text{z}}\approx 0$ as compared to Fig.~\ref{figSemiclassicalModel}(c). Intriguingly, the collective QFI is different from its semiclassical counterpart as it features a maximum at $ B_{\text{z}} = 0$, ensuring the validity of the QCRB. For this reason, we do not carry out an analysis of the corresponding four-level system.

Figure~\ref{figOverview}(d) depicts $\mathcal S_{B_{\text{z}   }}^{(n_{-} )}$ and $\mathcal I_{B_{\text{z}}}^{(Q)}$   as a function of the atom number for $B_{\text{z}}=0$ . As explained in Appendix~\ref{sec:collective:exact}, the SNR can be explicetly evaluated for $B_{\text{z}}=0$, where the signal is equal to the semiclassical result in Eq.~\eqref{eq:signal}, and the noise is given by 
\begin{eqnarray}
\left< \Delta \hat n_{- } ^2 \right>  &=&\overline n_+ + \frac{ \mathcal C_{A}^2}{4} \left(  \frac{ N \kappa_{\text{P}}  }{h_{\text{x}}^2 + \frac{\kappa_{\text{P}}^2}{4}  } 
+    \frac{N^2\tilde \gamma_{\text{z }}  h_x^2 } {\left( h_{\text{x}}^2 +\frac{\kappa_{\text{P }}^2 }{ 4 }\right)^2} \right)\tau \nonumber .\\
\label{eq:quantumNoiseCollective}
\end{eqnarray}
Likewise, the QFI explicitly reads
\begin{eqnarray}
\mathcal I_{B_{\text{z}}}^{(Q)}
&=&   \mu^2\left(   \frac{N \kappa_{\text{P}}  }{h_{\text{x}}^2 + \frac{\kappa_{\text{P}}^2}{4}  } 
+   \frac{N^2\tilde \gamma_{\text{z }}  h_x^2 } {\left( h_{\text{x}}^2 +\frac{\kappa_{\text{P }}^2 }{ 4 }\right)^2}\right)\tau.
\label{eq:quantumFisherInfoCollective}
\end{eqnarray}
 For small $N$, the QFI increases linearly and leaks through the pumping action $\propto \kappa_{\text{P}}$ to the environment. The SNR increases quadratically, as  the measurement noise is dominated by the photon-shot noise. 
 {\color{\markColorTwo} Once the atom number exceeds $N \approx10^{10}$, the behavior changes  as the second term  in the QFI starts to dominate. Now $\mathcal I_{B_{\text{z}}}^{(Q)}$ increases quadratically, i.e., it exhibits Heisenberg scaling $\propto N^2$, which originates from  the collective measurement backaction  $\propto \tilde  \gamma_{\text{z}}$ in Eq.~\eqref{eq:collectiveMasterEquation}.}
 However, the SNR becomes constant as the increasing dissipation  $\propto\tilde \gamma_{\text{z}}  $ leads to a quadratic scaling of $\langle \Delta \hat n_{-}^2 \rangle \propto \mathcal I_{B_{\text{z}}}^{(Q)}$ (for large $N$). Thus, besides the Heisenberg scaling of the QFI, the chosen measurement protocol fails to extract it because of the time integration of the intensity, which leads to a loss of information.  Crucially, the QFI as a function of the pump strength $\kappa_{\text{P} } $ in Fig.~\ref{figAnalysisCollective}(d) always exceeds the SNR of the cumulative statistics, verifying the microscopic consistency  in the collective model. 

{
  \color{\markColorTwo}  To physically interpret the origin of the Heisenberg scaling, we compare the collective model to the independent atom model in Sec.~\ref{sec:semiclassicalModel}. The lack of inter-atomic correlations in the semiclassical model  prevents --- by principle --- any super-linear scaling of the QFI in Eq.~\eqref{sec:quantumFisherInfoAtomSC}. In contrast, the collective model explicitly takes into account such correlations. According to the master equation in Eq.~\eqref{eq:collectiveMasterEquation}, the measurement-induced dissipation $\propto \gamma_z$ acts  a dissipative (i.e., non-unitary) non-linearity, which has a similar effect as a Hamiltonian collective  interaction between atoms. As such, it can create non-linear contributions to the time-integrated uncertainty in Eq.~\eqref{sec:quantumFisherInfoAtom}, and thus, to the QFI.
}

 { \color{\markColorTwo} Finally, we  motivate the scaling assumptions of  $\gamma_{\text{P}} =\kappa_{\text{P}}/N$ and  $\gamma =\tilde  \gamma/N\approx 0$, which are central for the interpretation of the results, by stating the following three reasons:
 	(i) As the atomic cloud is extended in z direction and confined in x direction, a probe photon cannot distinquish with which atom it has interacted. This indistinguishably gives rise to the collective dephasing term in Eq.~\eqref{eq:collectiveMasterEquation}. In contrast, the pump field (propagating in x direction) typically stimulates a Raman process in which a spontaneous photon is emitted. The origin of this photon could be --- in principle --- located, which destroys the collective action of the pump field. In our model, we have mimicked this effect by a scaled pumping rate. A similar consideration applies to $\gamma$, as the origin of a spontaneously emitted photon could be detected;
 	(ii) The pump rate is an external control parameter, and can be thus arbitrarily modified to follow a desired scaling;
 	(iii) The scaling of $\gamma_{\text{P}} $ and $ \gamma $ has been deliberately chosen such that the dynamics of the collective model corresponds to the semiclassical one in the absence of the measurement induced dissipation (i.e., for a small atom number). The agreement can be observed in Fig.~\ref{figAnalysisCollective}(a) and by comparing the term linear in $N$ in Eqs.~\eqref{eq:semiclassicalVariance} and \eqref{eq:quantumNoiseCollective} for $B_{\text{z}} =0$;  Other simplifying assumption	such as absorption limitations and technical noise are discussed in Appendix~\ref{sec:experimentalLimitations}.	
 }

\section{Discussion}

\label{sec:discussion}

 We have shown that a semiclassical description of optical magnetometers can violate the QCRB by orders of magnitude in the weak-dissipation, large-atom-number regime, revealing the fundamental insufficiency of models that assume independent atoms. This violation stems from the neglect of measurement-induced quantum correlations, which are essential to uphold the QCRB. A collective spin model, which microscopically incorporates these correlations, fully respects the QCRB. Notably, this collective description predicts Heisenberg scaling of the QFI with atom number. 

Crucially, the emergent Heisenberg scaling is not driven by direct interatomic interactions or initial entanglement, but by correlations generated dynamically through the continuous measurement process itself, sustained even in a dissipative stationary state. This establishes a new paradigm for quantum sensing in which measurement-backaction, rather than Hamiltonian dynamics, provides the quantum advantage in a macroscopic ensemble. 
{\color{\markColorTwo}  While our research identifies the measurement backaction-induced correlations as a quantum resource to reach the Heisenberg scaling of the QFI,  time-integrated measurements as considered here are unable to \textit{read out} the quantum information. We envision two fundamentally distinct approaches to read out the QFI: (i) By conditioning the time evolution of the collective model on the measurement record in a Bayesian way, one can construct more sensitive estimators than the time-averaged protocol investigated here~\cite{Kuzmich2000,Gammelmark2013,Barberena2024,AmorosBinefa2025}. Doing so also allows to analyze the information gain adhering to the transition from the unconditional to the conditional dynamics; (ii) Alternatively, the QFI may be stored in a complex state involving entanglement of the  atomic ensemble and the probe field. In this case, a Bayesian analysis will likely not be capable of extracting the QFI. Instead, more sophisticated approaches using quantum backaction evading measurements or the closely-related quantum-decoding protocols will be required~\cite{Moeller2017,Tsang2012,Yang2023}. These  sensing protocols will be investigated in future research.
}

A key experimental signature distinguishing the two models can be formulated using the QCRB. As the  QFI fulfills $\mathcal I_{B_{\text{z}}  }^{(Q)}   = 4\mu^2 	\left( \left< \Delta \hat n_{- }^2  \right> -\overline n_{+}\right)/\mathcal C_{A}^2$ according to Eqs.~\eqref{sec:quantumFisherInfoAtom} and \eqref{eq:collectiveNoise}, the QCRB in Eq.~\eqref{eq:FisherInfo-SNR} results in the  consistency condition:
\begin{equation}
  \left< \Delta \hat n_{- } ^2 \right>  > \frac{\Omega^2}{2\mu \epsilon_{\Delta} } \left<\partial_{B_{\text{z}}} \hat n_{-}  \right>.
 \label{eq:experimentalConsistencyCondition}
\end{equation}
 The semiclassical model can violate Eq.\eqref{eq:experimentalConsistencyCondition} for large atom numbers, whereas the collective model respects it. An experimental test of this inequality can therefore serve as a direct probe for the presence of macroscopic quantum correlations, while a violation would challenge the quantum description of light–matter interaction in a macroscopic quantum system.

The  different predictions of the two models—regarding polarization rotation, noise magnitude, and the validity of Eq.~\eqref{eq:experimentalConsistencyCondition} provide a clear blueprint for their experimental discrimination. Such experiments would test the quantum–classical boundary in a system of up to $10^{13} $ atoms. The predicted, inherently robust Heisenberg scaling opens new pathways for achieving quantum-enhanced sensitivity in practical, room-temperature sensors without requiring precise state preparation or  atomic interactions.

	\begin{acknowledgments}
		G.E. acknowledges NSFC Grant No. W2432004. X.W. acknowledges NSFC Grant No. 12404409 and SUSTech Presidential Postdoctoral Fellowship. J.Y.L. acknowledges NSFC Grant No. 11774311. J.F.C. acknowledges NSFC Grants No. 92476102, No. 92265109, and Guangdong Key Project Grant No. 2022B1515020096.\\
	\end{acknowledgments}

Numerical data and corresponding source code are openly available~\cite{Engelhardt2025source}.

\appendix

\section{Hamiltonian}

\label{sec:Hamiltonian}

 We deploy the common input-output representation to describe the light-matter Hamiltonian in  an interaction picture defined by the unperturbed matter and photonic systems, such that the Hamiltonian in Eq.~\eqref{eq:microscopicHamiltonian} can be represented as
\begin{eqnarray}
\hat H(t) &=& \sum_{\xi =(\eta ,t_\xi)} \sum_m  G_{\xi,m}(t) \left[ \hat V_{\xi,m} (t)  \hat a_\xi^\dagger  + \text{h.c.} \right] , \nonumber \\
&=& \sum_{\xi =(\eta ,t_\xi)} \hat H_{\xi}(t)
\label{eq:hamiltonian}
\end{eqnarray}
where $G_{\xi,m}(t) = d_A \sqrt{\hbar \omega_{\text{p}}dt /2\epsilon \mathcal A  } \delta(t - t_{\xi} - r_m/c ) \propto dt^{-1/2} $ is a time-local light-matter interaction parameter whose scaling $\propto dt^{-1/2}$ is important for the consistent derivation of the master equation. The position of the atoms along the z direction is denoted by  $r_m$. The photon operators $\hat a_\xi,\hat a_\xi^\dagger $ quantize the probe laser, where $\xi = (\eta, t_{\xi})$ labels the polarization direction $\eta=\text{1},\text{2}$ as well as the time $t_\xi$ at which the photonic operator $\hat a_\xi $ interacts with an atom at $r_m=0$. Thereby, the photonic operators can be imagined to quantize a train of laser pulses propagating in the positive z direction, which couple to the matter operator 
\begin{equation}
\hat V_{\xi,m}(t) = \sum_{a=\pm 1} e^{i a(\eta-1) \frac{\pi}{2} }  \left| g_{m,a}\right> \left< e_{m,-a}  \right|  e^{i\epsilon_{\Delta} t}  .
\label{eq:interactionPictureHamiltonian}
\end{equation}
The quantized Hamiltonian in Eq.~\eqref{eq:hamiltonian} represents thus the system in the photonic basis $\hat a_\xi ,\hat a_{\xi}^\dagger$, in which the probe field in measured.

{
\color{ \markColorOne}
The initial condition of the light-matter interaction is given by
\begin{eqnarray}
\rho_{\text{tot} }( 0)  = \rho_{\text{M} }( 0) \otimes  \prod_{\xi }  \left| \alpha_\xi \right> \left< \alpha_\xi \right|.
\label{eq:intialState}
\end{eqnarray}
Thereby,  $\rho_{\text{M} }( 0)$ is the matter density matrix, and $\left| \alpha_\xi \right>$ denotes a photonic coherent state of  mode $\hat a_\xi$, i.e., $\hat a_\xi \left| \alpha_\xi \right> =\alpha_\xi  \left| \alpha_\xi \right>$.  To obtain a well-defined scaling when setting up the master equation, we assume that  $\left| \alpha_{\xi} \right|^2 = \dot \nu_{\xi}   dt$ with a real-valued photon flux $\dot \nu_{\xi} = d\nu_{\xi}/dt $ attributed to the mode $\hat a_\xi$, where $dt$ denotes the time increment. 

}

\section{Measurement statistics}

\label{sec:fisherInformation}

Here, we review basic properties of the QFI for self consistency in Appendices~\ref{sec:modifiedCramerRaoBound} and \ref{sec:app:quantumFisherInformation}. In Appendix~\ref{sec:app:fullCountingStatistics}, we recall the basic concepts of full-counting statistics, which we use to calculate the measurement statistics.

\subsection{Modified Cram\'er-Rao bound}

\label{sec:modifiedCramerRaoBound}

Here, we derive the representation of the QFI in Eq.~\eqref{sec:quantumFisherInfoAtom} for Hamiltonian parameter estimation using an argument from linear-response theory. For a more general derivation, we refer to Ref.~\cite{Gorecki2025}. Thereby, we assume that the Hamiltonian  depends on the parameter to be measured $X$, i.e., $\hat H = \hat H_{X } $.

Applying the Heisenberg inequality  to an arbitrary operator $\hat A $ and the operator
\begin{eqnarray}
\hat  {\mathcal F}=  \int_{0}^{\tau}   \partial_{X} \hat H_{X} (t)  dt ,
\end{eqnarray}
we obtain
\begin{eqnarray}
\left< \Delta \hat {\mathcal F} ^2 \right> \left< \Delta \hat A ^2 \right>
&\equiv&\left< \left( \hat {\mathcal F} -\left<\hat{\mathcal F} \right>\right)^2 \right> \left< \left( \hat A - \left<  \hat A \right>\right)^2 \right>\nonumber \\
&\ge& \frac{1}{4}  \left| \left< \left[ \hat {\mathcal F} , \hat A \right] \right>\right|^2 \nonumber \\
&= &\frac{1}{4}  \left| \int_{0}^{\tau} dt   \left< \left[   \partial_{X} \hat H_{X} (t)   , \hat A \right] \right>\right|^2, \nonumber \\
\label{eq:heisenbergInequality}
\end{eqnarray}
where the expectation value refers to the initial state of the joint light-matter state.
Within linear-response theory, we can identify 
\begin{equation}
\int_{0}^{\tau} dt   \left< -i\left[   \partial_{X} \hat H_{X} (t)   , \hat A \right] \right> = \left<\frac{d\hat A}{dX} \right> 
\end{equation}
as the response of the observable $\hat A$ for a small perturbation of the Hamiltonian in $X$.
Convincing ourselves that
\begin{eqnarray}
\mathcal I_X^{(Q)}  &=& 4  \left< \left( \hat  {\mathcal F}  -\left<\hat  {\mathcal F} \right>\right)^2 \right>, \nonumber \\
&=& 4\int_{0}^{\tau} dt_1\int_{0}^{\tau}dt_2  \left<  \Delta \hat F(t_1) \Delta \hat F(t_2) \right> ,
\label{eq:definitionFisherInfo}
\end{eqnarray}
where $ \mathcal I_X^{(Q)}$ is the QFI defined in Eq.~\eqref{sec:quantumFisherInfoAtom}, we obtain the inequality  
\begin{equation}
\mathcal I_X^{(Q)} \ge  \frac{\left<  \partial_{X}\hat A   \right>^2 }{\left< \Delta \hat A^2  \right> }
\label{eq:CramerRaoBound},
\end{equation}
which is an equivalent version of the celebrated QCRB. 
Noteworthy, Eq.~\eqref{eq:CramerRaoBound} is also valid for continuous measurements (such as in optical magnetometry), as a continuous measurement in time can be considered an instantaneous measurement in position space.

\subsection{Quantum Fisher information}

\label{sec:app:quantumFisherInformation}

Here we provide  alternative expressions for the QFI in Eq.~\eqref{eq:definitionFisherInfo} which are easier to evaluate. The time-evolved   wave function at time $\tau$ reads
\begin{equation}
\left|\Psi_X  \right>  = e^{-i \hat H_{X} \tau} \left|\Psi(0) \right> ,
\label{eq:parameterDependentState}
\end{equation}
where $\left|\Psi(0) \right>$ denotes the initial state. We now show that the QFI can be  expressed as~\cite{Gammelmark2014}
\begin{equation}
\mathcal I_{X}^{(Q)}   = 4 \left[\left<\partial_X \Psi_X  \mid  \partial_X \Psi_X\right>  - \left|\left< \Psi_X  \mid  \partial_X \Psi_X\right> \right|^2 \right].
\label{eq:fisherInformationQuantuGeometry}
\end{equation}
Without loss of generality, we consider the QFI at $X=0$. To this end, we expand
\begin{equation}
\hat H_{X}   =  \hat H_{0}  + \hat FX +\mathcal O\left[ X^2\right],
\end{equation}
where $\hat F  =\partial_X \hat H_{X=0} $. The time-evolution operator in Eq.~\eqref{eq:parameterDependentState} can be expressed as 
\begin{eqnarray}
e^{-i \hat H_{X} t}   = e^{-i \hat H_{0} t} \hat  U_{X}(t)  ,
\end{eqnarray}
where $\hat  U_{X}(t) $ is the time-evolution operator in the interaction picture defined by $\hat H_{X=0}$. Expanding in orders of $X$, it reads
\begin{equation}
\hat U_{X}(t)   =  1 -i X \int_{0}^{\tau}dt \hat F(t) +\mathcal O(X^2),
\end{equation}
where the time dependence of $\hat F(t)$ is determined by $\hat H_0$.
Using this expression to evaluate Eq.~\eqref{eq:fisherInformationQuantuGeometry}, we directly find the integral form  of the QFI in Eq.~\eqref{eq:definitionFisherInfo}.

Alternatively, the QFI in Eq.~\eqref{eq:fisherInformationQuantuGeometry} can be expressed as
\begin{eqnarray}
\mathcal I_{X}^{(Q)}   &=& -  \partial^2_\delta   \log \text{tr} \left[ \rho_{\delta}(\tau)   \right]_{\delta=0} ,\nonumber\\  \nonumber\\
\rho_{\delta}(\tau)   &=& \left|\Psi_{\delta} \right> \left<   \Psi_{-\delta} \right|,
\label{eq:FisherInfo:alternative}
\end{eqnarray}
which is a modified version of the ones in Refs.~\cite{Ilias2022,Yang2023}. This expression is suitable for the evaluation in open quantum systems, where the time-dependence of the  density matrix $ \rho_{\delta} $ can be obtained by integrating a generalized master equation
\begin{eqnarray}
\frac{d}{dt}  \rho_{\delta} &=&  \mathcal L_{ \delta } \rho_{\delta}  \nonumber \\
&=&  -i \left[ H_{\text{M},\delta}  \rho_{\delta}  -  \rho_{\delta}H_{\text{M},-\delta} \right] +\mathcal L^{\text{D} }  \rho_{\delta}, \label{eq:generlizedMasterEquationFisher}
\end{eqnarray}
as can be seen by comparing Eqs.~\eqref{eq:fisherInformationQuantuGeometry} and~\eqref{eq:parameterDependentState}. Thereby, we assume that only the matter contribution of the  Hamiltonian $H_{\text{M},X} $ parametrically depends on $X$, while the terms giving rise to dissipation (described by $\mathcal L^{\text{D} } $)  are independent of $X$.

In the long-time limit, the QFI can be efficiently evaluated using Eq.~\eqref{eq:generlizedMasterEquationFisher}~\cite{Gammelmark2014,Ilias2022}. Let us denote the eigenvalues and eigenvectors of the  Liouvillian $\mathcal L_{\delta  } $  by  $\lambda_{a,\delta }$ and  $\rho_{a,\delta }$. Labeling the eigenvectors such that $\lambda_{a=0,\delta }$ is $\lambda_{a=0,\delta  =0}=0$, we find in the long-time limit
\begin{equation}
\lim_{t\rightarrow \infty }\rho_{\delta  } (t)  =  e^{\lambda_{0,\delta } t  }  \rho_{0,\delta },
\end{equation}
i.e., the generalized density matrix is dominated by the $a=0$ eigenvalue. Using now Eq.~\eqref{eq:FisherInfo:alternative},  we find the asymptotic behavior of the QFI
\begin{equation}
\mathcal I_{X}^{(Q)}  (t) = -\partial^2_\delta  \left. \lambda_{0,\delta  }  \right|_{\delta  =0} t ,
\label{eq:fisherInformationLongTime}
\end{equation}
which can be efficiently evaluated numerically.

\subsection{Full-counting statistics}
\label{sec:app:fullCountingStatistics}

Here, we introduce the basic methodology which we use to calculate the measurement statistics of  operators $\hat A$ in Eq.~\eqref{eq:CramerRaoBound}.
We are interested in the measurement statistics of the set of photonic occupation operators $\hat n_{\xi} = \hat a_{\xi}^\dagger \hat a_{\xi} $, which are closely related to the intensity operator defined as
\begin{equation}
\hat{\mathcal I}_\eta(t) \equiv  \frac{1}{dt \mathcal A} \hat a_{\xi}^\dagger \hat a_{\xi},
\label{eq:def:intensityOperator}
\end{equation}
where  $dt$ is the time increment corresponding to the operators $\hat a_{\xi}^\dagger , \hat a_{\xi}$. Using $\hat{\mathcal I}_\eta(t) $ we define the integrated intensity operator in Eq.~\eqref{def:photonNumbers}.

The corresponding photon numbers are denoted by $n_{\xi}$. In terms of the  photonic probabilities $p_{\boldsymbol n}$, where  $\boldsymbol n  =  \left( n_{1}  ,\dots,n_{\xi_{\text{max} }}  \right)$ is a vector of photon numbers, the moment- and cumulant-generating functions are defined as
\begin{eqnarray}
M_{\boldsymbol \chi}  &\equiv & \sum_{\boldsymbol n} p_{\boldsymbol n} e^{-i\boldsymbol  \chi \cdot \boldsymbol n } 
= \text{tr} \left[ e^{-i \boldsymbol \chi \cdot \hat {\boldsymbol n } }   \rho_{\text{tot} }  \right] ,\nonumber \\
K_{\boldsymbol \chi} &\equiv &  \log M_{\boldsymbol \chi}  ,
\label{eq:def:momentGenFct} 
\end{eqnarray}
where $\rho_{\text{tot} }$ denotes the density matrix of the full system, and $\text{tr}\left[ \bullet\right]$ is the trace over all degrees of freedom. The parameters $\boldsymbol \chi = \left(\chi_1 , \dots ,\chi_{\xi_{\text{max}  }}  \right)$  are the so-called counting fields conjugated to the photon numbers. 

The derivatives of the cumulant-generating function  with respect to the counting fields generate the cumulants of the probability distribution. For instance, the first- and second-order  cumulants,
\begin{eqnarray}
\kappa_1^{(\xi)}  &\equiv &  i\partial_{\chi_\xi} K_{\boldsymbol \chi =\boldsymbol 0}  =  \left< \hat n_\xi(t)\right>  = \overline n_{k},\nonumber\\
\kappa_2^{(\xi)}   &\equiv &-\partial_{\chi_\xi}^2 K_{\boldsymbol \chi =\boldsymbol 0}  =   \left< \hat n_\xi^2\right> -\left< \hat n_k\right>^2 ,
\label{eq:lowOrderCumulants}
\end{eqnarray}
are the mean and the variance of the photonic probability distribution. Likewise, mixed derivatives produce covariances of the probability distribution.  Higher-order cumulants provide information about the deviation of the probability distribution from a Gaussian distribution.

As explained in Eq.~\eqref{def:photonNumbers}, we are interested in time-integrated measurements, which distinquish only between the polarization degrees. Inspection of the moment-generation in Eq.~\eqref{eq:def:momentGenFct} reveals, that this can be simply achieved by setting
\begin{equation}
\chi_ \xi = \chi_{\eta},
\label{eq:timeInvariantCountingField}
\end{equation}
that is, neglecting the time dependence of the counting-field label.

\section{Semiclassical model}

\label{sec:app:semiclassicalModel}

Because of the macroscopic number of quantum-emitters, it is not possible to treat the system in a fully microscopic fashion. To obtain  quantitative predictions, we develop here a hybrid semiclassical-quantum-trajectory approach  based on  successive light-atom interactions. The quantum trajectory approach developed in Appendix~\ref{sec:masterEquation} ensures that the light-matter interaction for each individual atom is on a microscopic footing, and thus consistent with the QCRB. The semiclassical amendment, which is introduced in Appendices~\ref{sec:flowEquations} and \ref{sec:polarizationMeasurements}, deploys a stochastic mean field to deal with the ensemble of atoms. In Appendices~\ref{sec:fourLevelSystemSemiclassical}-\ref{sec:analyticalCalculations}, we apply the formalism to the system in Appendix~\ref{sec:Hamiltonian}. Finally  in Appendix~\ref{sec:cumulantsIntegralForm}, we  derive the relation between full-counting statistics and integral expressions given in the main text.

\subsection{Generalized master equation}

\label{sec:masterEquation}

In this appendix, we derive an expression for the moment-generating function of the photonic field after interaction with a single atom $m$ located at $ r_m =0$ which is on an equal footing with a quantum trajectory approach~\cite{Wiseman1993,Wiseman2010,Landi2024}.  For brevity, we suppress the index $m$ in the following.

After interaction of the radiation field with the matter system, the moment-generating function defined in Eq.~\eqref{eq:def:momentGenFct}  is given by
\begin{eqnarray}
M_{\boldsymbol \chi }(\tau)  &=&  \text{tr} \left[ e^{-i \boldsymbol \chi \cdot  \hat{ \boldsymbol n } } \tilde U (0, \tau )  \rho_{\text{tot}}(0) \tilde U^\dagger (0, \tau)  \right] ,
\end{eqnarray}
where $ \tilde U (0, \tau )$ denotes the time-evolution operator corresponding to Eq.~\eqref{eq:hamiltonian}, and the initial state of the light-matter interaction $\rho_{\text{tot}}(0)$ is in Eq.~\eqref{eq:intialState}.

For small time-increments $dt$, the time-evolution operator $\tilde U $  can be expanded as
\begin{eqnarray}
\tilde U (t,t+dt)&=&  1 -i   dt  \sum_{\xi}  \hat H_{\xi}(t)      \nonumber \\
&-&   \frac{1}{2} dt^2 \sum_{\xi_1,\xi_2}   \hat H_{\xi_1}(t)   \hat H_{ \xi_2}(t) \nonumber \\
&+& \mathcal O \left[ G_{\xi}^3(t) \right].
\label{eq:timeEvolutionIncrement}
\end{eqnarray}
As  the light-matter coupling parameters $ G_{\xi}(t)$ interact with the quantum system only in a particular instant of time, we can  neglect higher expansion orders within the Markov assumption in the quantum trajectory approach.  

Let us define now the generalized density matrix of the matter system at time $t$  via the relation
\begin{eqnarray}
\tilde \rho_{\boldsymbol \chi}(t ) &=&   \text{tr}_{\text{L}} \left[ e^{-i \sum_{t^\prime\leq t}\boldsymbol \chi_{t^\prime} \cdot \hat{\boldsymbol n}  } \tilde U (0, t )  \rho_{\text{tot}}(0) \tilde U^\dagger ( 0,t )  \right], \nonumber\\
\label{eq:reducedDensityMarix}
\end{eqnarray}
where $ \text{tr}_{\text{L}}\left[\bullet \right]$ denotes the trace over the photonic degrees of freedom. We have defined $\boldsymbol \chi_{t}$ as the vector of counting fields whose corresponding photonic operators interact with the matter system at time $t$.

The next step is to express $\tilde \rho_{\boldsymbol \chi}(t +dt )$ in terms of   $\tilde \rho_{\boldsymbol \chi}(t )$ and  the time-evolution operator in Eq.~\eqref{eq:timeEvolutionIncrement}. Let us denote  the set of photonic modes, which are going to interact with the matter system during the time interval $(t,t+dt)$ by $\Lambda_t$. Then we can evaluate
\begin{widetext}
	\begin{eqnarray}
	\tilde \rho_{\boldsymbol \chi} (t+dt)&\equiv& \text{tr}_{\text{L}}\left[e^{-i \boldsymbol \chi_{t+dt} \cdot \boldsymbol n  }  U(t,t+dt) \tilde \rho_{\boldsymbol \chi}(t)  \prod_{\xi\in \Lambda_t } \left|\alpha_\xi \right> \left< \alpha_\xi \right|  U^\dagger(t,t+dt) \right] \nonumber \\
	&=&  \tilde \rho_{\boldsymbol \chi} M_{\boldsymbol \chi, \Lambda_t  }(0) \nonumber\\
	&-&i  dt\sum_{\xi\in \Lambda_t}\left[G_{\xi}(t )  \left( \alpha_{\xi}^*  \hat V_{\xi} (t)   e^{-i\chi_{\xi} }  + \alpha_{\xi}  \hat V_{\xi}^{\dagger}(t)   \right)   \tilde   \rho_{\boldsymbol \chi}    
	+ i  \tilde \rho_{\boldsymbol \chi}   G_{\xi}(t) \left( \alpha_{\xi}^*    \hat V_{\xi} (t)    + \alpha_{\xi}  \hat V_{\xi}^{\dagger}(t)  e^{-i\chi_{\xi} }  \right)    \right]  M_{\boldsymbol \chi, \Lambda_t  }(0)  \nonumber \\
	&+&  dt^2 \sum_{\xi\in \Lambda_t} \left[ e^{-i\chi_{\xi} }     G_{\xi}^2(t)   \hat V_{\xi}(t)   \tilde  \rho_{\boldsymbol \chi}    \hat V_{\xi}^{\dagger}(t) 
	-    \frac{1}{2}  G_{\xi}^2(t)    \left \lbrace \hat V_{\xi}^{\dagger}(t)  \hat V_{m,\xi}(t)  ,  \tilde  \rho_{\boldsymbol \chi}    \right \rbrace \right] M_{\boldsymbol \chi, \Lambda_t  }(0))  \nonumber \\
	&+& \mathcal O \left[ G_{\xi}^3(t) \right] ,
	\label{eq:generlizedDensityMarrixTimeIncrement}
	\end{eqnarray}
	
\end{widetext}
where we have utilized
\begin{eqnarray}
\hat a_\xi  e^{-i \chi_\xi  \hat n_\xi  }   =   e^{-i \chi_\xi   \hat n_\xi  } \hat a_\xi e^{-i\chi_\xi}
\end{eqnarray}
and 
\begin{equation}
\left< \alpha_\xi \left| e^{-i \chi_\xi  \cdot \hat n_\xi  } \right| \alpha_\xi \right> = \exp \left[ \left(e^{-i\chi_{\xi} } -1 \right)\left|\alpha_\xi \right|^2  \right],
\end{equation}
which is the moment-generating function of the Poisson distribution. These terms give rise to the initial moment-generating function
\begin{eqnarray}
M_{\boldsymbol \chi, \Lambda_t  }(0) &=&  \prod_{\xi\in \Lambda_t }  \exp \left[ \left(e^{-i\chi_{\xi} } -1 \right)\nu_{\xi} \right],
\end{eqnarray}
of the modes $\xi \in \Lambda_t $, where we have introduced $\nu_{\xi}= \left|\alpha_\xi \right|^2$.

Using Eq.~\eqref{eq:generlizedDensityMarrixTimeIncrement}, we can derive an expression for the final moment-generating function
\begin{eqnarray}
M_{\boldsymbol \chi\mid \boldsymbol \nu }(\tau)  &=&  M_{\text{dy},\boldsymbol \chi\mid \boldsymbol \nu  }(\tau)  M_{\boldsymbol \chi\mid \boldsymbol \nu}(0),
\label{eq:momentGenFctProduct}
\end{eqnarray}
where
\begin{eqnarray}
M_{\boldsymbol \chi\mid \boldsymbol \nu}(0) &=&  \prod_{\xi }  \exp \left[ \left(e^{-i\chi_{\xi} } -1 \right)\nu_{\xi} \right].
\label{eq:initialMomentGenFunction}
\end{eqnarray}
The term $M_{\text{dy},\boldsymbol \chi\mid \boldsymbol \nu  }(\tau)$ is denoted as the dynamical moment-generating function and is given by 
\begin{eqnarray}
M_{\text{dy},\boldsymbol \chi\mid \boldsymbol \nu  }(\tau)   &=& \text{tr} \left[ \rho_{\boldsymbol \chi   } (\tau) \right]  ,
\label{eq:dynamicalMomentGenFkt}
\end{eqnarray}
where $\rho_{\boldsymbol \chi}$ is the reduced density matrix. It can be obtained by the solution of the generalized master equation
\begin{eqnarray}
\frac{d}{dt}\rho_{\boldsymbol \chi}    &=&   -i \left[  \hat {\mathcal H}_{\boldsymbol \chi}(t)  \rho_{\boldsymbol \chi}   - \rho_{\boldsymbol \chi}  \hat {\mathcal H }_{-\boldsymbol \chi}^\dagger(t)    \right] \nonumber  \\
&+&\mathcal L_{\boldsymbol \chi}^{\text{D}}(t)\rho_{\boldsymbol \chi} 
+\mathcal L^{ \text{D}}\rho_{\boldsymbol \chi} ,
\label{eq:generalizedMasterEquation}
\end{eqnarray}
where the non-Hermitian Hamiltonian is given by
\begin{eqnarray}
\hat {\mathcal H}_{\boldsymbol \chi}(t)   =  \sum_{\xi }G_{\xi}(t) \left(  \alpha_{\xi}^*   \hat V_{\xi} (t)   e^{-i\chi_{\xi} }  +\alpha_{\xi}  \hat  V_{\xi}^{\dagger} (t)   \right).  \nonumber\\
\end{eqnarray}
Importantly, the non-Hermitian character of the generalized master equation encodes the counting statistics of the photons. The counting-field-dependent dissipation terms in Eq.~\eqref{eq:generalizedMasterEquation} are given by
\begin{equation}
\mathcal L_{\boldsymbol \chi}^{\text{D}}(t)\rho	=\sum_\xi dt G_{\xi}^2(t)  D_{\chi_\xi} [\hat V_{\xi} (t) ]\rho,\nonumber  
\end{equation}
where
\begin{equation}
D_\chi [\hat O]\rho   = e^{-i\chi} \hat O \rho \hat O^\dagger +\frac{1}{2} \left\lbrace  \hat O^\dagger \hat O, \rho \right\rbrace
\label{eq:generalizedDissipator}
\end{equation}
is a counting-field-dependent dissipator. Using the moment-generating function in Eq.~\eqref{eq:momentGenFctProduct}, we can calculate the cumulants of the photonic probability distribution using Eq.~\eqref{eq:lowOrderCumulants}. To relate $G_{\xi}(t) $ to physical quantities, we introduce
\begin{eqnarray}
\Omega_{\xi} &=&  2 G_{\xi}(t) \alpha_{\xi} ,\nonumber \\
\gamma_{\text{z}} &=&  dt G_{\xi}^2(t) 
\end{eqnarray}
as the Rabi frequency and spontaneous emission rate in positive z direction. Reminding ourselves that $\left| \alpha_{\xi} \right|^2 = \dot \nu_{\xi}   dt$ with photon flux $\dot \nu_{\xi} = d\nu_{\xi}/dt $, we find the relation
\begin{equation}
\gamma_{\text{z}} = \frac{ \left| \Omega_{\xi}\right|^2 }{4\dot \nu_{\xi}},
\label{eq:masterEquationConsistency}
\end{equation}
which must be fulfilled to guarantee microscopic consistency of the generalized master equation in Eq.~\eqref{eq:generalizedMasterEquation}. In experiments, the Rabi frequency and the photon flux can be easily determined, from which we can then obtain a consistent $\gamma_{\text{z}}$.

In the semiclassical treatment of spectroscopy, the spontaneous emission rate $\gamma_{\text{z}}$ is extremely small, such that the  number of spontaneously emitted photons  in the z direction is negligible. However, these terms can  significantly influence the dynamics in the collective model.  Phenomenologically, we have added another dissipation term $\mathcal L^{\text{D}}\rho_{\boldsymbol \chi}$, which accounts for spontaneous emission not along the z direction, or other optical pumping terms.

In the following, we consider a  time-invariant photonic field and a time-integrated measurement statistics. For this reason, we  replace $\xi =(\eta ,t_\xi)$ by $\eta$ in the generalized master equation Eq.~\eqref{eq:generalizedMasterEquation}. See also the explanations for Eq.~\eqref{eq:timeInvariantCountingField} in this context. In doing so, the asymptotic moment-generating function is given by
\begin{equation}
M_{\text{dy},\boldsymbol \chi\mid \boldsymbol \nu  }(\tau)  \rightarrow e^{\lambda_{0,\boldsymbol \chi} \tau} ,
\label{eq:symptoticMomentGeneratingFunction}
\end{equation}
where $\lambda_{0,\boldsymbol \chi}$ is the eigenvalue with the largest real part of the Liouvillian in Eq.~\eqref{eq:generalizedMasterEquation}. This is analog to the calculation of the QFI in Eq.~\eqref{eq:fisherInformationLongTime}.

\subsection{Semiclassical flow equations}

\label{sec:flowEquations}

In Sec.~\ref{sec:masterEquation}, we have derived the moment-generating function for the light-matter interaction with a single atom. To deal with a macroscopic ensemble of atoms, we develop here a semiclassical approach. Thereby, we consider   \textit{stochastic} photonic mean fields $\nu_{\eta}  = \left|\alpha_{\eta}\right|^2 $, where $\alpha_{\eta}$ describes the coherent photonic state  in Eq.~\eqref{eq:intialState} with polarization direction $\eta \in \left \lbrace\text{1,2} \right \rbrace $.  The following treatment focuses on the first- and second-order cumulants, which we  investigate in this work.

The stochastic mean fields are described by the  probability distribution  $p_{\boldsymbol \nu }(z)$, where $z$ refers here to a position within the atomic cloud. During the propagation of the photonic field through the atomic ensemble, the probability distribution will be continuously updated according to the light-matter interaction. In the following, we derive and analyze flow equations for the low-order cumulants of  $p_{\boldsymbol \nu }(z)$. Crucially, we thus distinguish between the photonic probability distribution $p_{\boldsymbol n }(z)$ and the stochastic mean-field probability distribution $p_{\boldsymbol \nu }(z)$.

We derive the flow equations in terms of the averaged moment-generating function defined as
\begin{eqnarray}
\overline M_{\boldsymbol \chi}(z) &=& \int d\boldsymbol \nu p_{\boldsymbol \nu }(z) M_{\boldsymbol \chi\mid \boldsymbol \nu }(z),
\label{eq:averagedMomentGeneratingFunction}
\end{eqnarray}
where $ M_{\boldsymbol \chi\mid \boldsymbol \nu }(z) $ is the moment-generating function for the photonic probability distribution  in Eq.~\eqref{eq:momentGenFctProduct} conditioned on a specific coherent mean-field value $ \boldsymbol \nu$. We now determine the updated probability distribution $p_{\boldsymbol \nu }(z+dz)$, such that we can represent the averaged moment-generating in Eq.~\eqref{eq:averagedMomentGeneratingFunction} in the form
\begin{eqnarray}
\overline M_{\boldsymbol \chi}(z) &\equiv& \int d\boldsymbol \nu p_{\boldsymbol \nu }(z+dz)e^{\boldsymbol \nu\cdot  (e^{i\boldsymbol \chi} -1 )},
\label{eq:averagedMomentGeneratingFunctionAlt}
\end{eqnarray}
where $e^{i\boldsymbol \chi}$ refers to a component-wise evaluation of the counting-field vector $\boldsymbol \chi$. The exponential factor in Eq.~\eqref{eq:averagedMomentGeneratingFunctionAlt} is the moment-generating function of coherent states with mean photon numbers $\boldsymbol \nu$.
In other words, we aim to find a mean-field distribution, such that the resulting ensemble of coherent photonic states gives rise to the same photonic probability distribution as Eq.~\eqref{eq:averagedMomentGeneratingFunction}.

As we are interested in describing the first two cumulants, we parameterize
\begin{eqnarray}
p_{\boldsymbol \nu }(z)  =  \frac{1}{ 2\pi \det \boldsymbol \Sigma_z} \exp \left[  -\frac{1}{2} \left(\boldsymbol  \nu -\overline{ \boldsymbol  \nu}_{z} \right) \boldsymbol  \Sigma_z^{-2} \left(\boldsymbol  \nu -\overline{ \boldsymbol  \nu}_{z} \right)  \right] ,\nonumber \\
\label{eq:gaussianProbabilityDistribution}
\end{eqnarray}
where $\overline{ \boldsymbol  \nu}_{z} $ denotes the vector of mean values of the stochastic mean field, and $\boldsymbol  \Sigma_z^{2} $ the corresponding covariance matrix.  
Moreover, when expanding the conditioned moment-generating function as
\begin{eqnarray}
M_{\boldsymbol \chi\mid \boldsymbol \nu }(z) &=& \exp \left( -i  \boldsymbol  \kappa_{1, \boldsymbol \nu }\cdot  \boldsymbol \chi - \frac{1}{2} \boldsymbol   \chi  \boldsymbol  \kappa_{2,\boldsymbol  \nu } \boldsymbol \chi   +\mathcal O( \boldsymbol \chi^3)  \right)\nonumber \\
&=&  \exp  \left[ -i  \boldsymbol \kappa_{1, \overline{ \boldsymbol  \nu}_z }   \cdot  \boldsymbol \chi - \frac{1}{2}  \boldsymbol  \chi   \boldsymbol \kappa_{2, \overline{\boldsymbol \nu}_z }  \boldsymbol \chi     \right. \nonumber \\
&& \left. -i \sum_{\eta = \text{1,2} }   \left( \partial_{\overline \nu_{\eta} }  \boldsymbol \kappa_{1, \overline{\nu}_z }  \cdot  \boldsymbol \chi  \right) \tilde \nu_{\eta} + \mathcal O (\tilde{ \boldsymbol \nu}^2 , \tilde {\boldsymbol\nu } \boldsymbol \chi^2, \boldsymbol \chi^3) \right], \nonumber \\
\label{es:conditioneMomentGenFunctionExpanded}
\end{eqnarray}
where $\boldsymbol \kappa_{1,  \boldsymbol  \nu_z }   $ ($\boldsymbol \kappa_{2,  \boldsymbol  \nu_z }   $) denotes the vector (matrix) of first (second) order  cumulants corresponding to the moment-generating function $M_{\boldsymbol \chi\mid \boldsymbol \nu }(z)$. In the second line, we have expanded the cumulants up to leading order in  $ \tilde{\boldsymbol  \nu} = \boldsymbol \nu  - \overline{ \boldsymbol  \nu}_z  $, as higher-order terms in the expansion do not influence the final flow equations.
Because of the product form of the conditioned moment-generating function in  Eq.~\eqref{eq:momentGenFctProduct}, we find
\begin{eqnarray}
\boldsymbol \kappa_{1, \boldsymbol \nu  }   &=&  \boldsymbol  \nu  + \boldsymbol  \kappa_{ \text{dy}, 1,\boldsymbol \nu   } , \nonumber \\
\boldsymbol \kappa_{2, \boldsymbol \nu}   &=&  \text{diag}\,  \boldsymbol  \nu   + \boldsymbol  \kappa_{ \text{dy}, 2, \boldsymbol \nu   },
\end{eqnarray}
where $ \text{diag}\,  \boldsymbol  \nu $ denote a diagonal matrix with diagonal elements given by $   \boldsymbol  \nu $ , and
\begin{eqnarray}
\left[  \boldsymbol  \kappa_{ \text{dy}, 1,\boldsymbol \nu   } \right]_\eta  &=&      i\partial_{\chi_{\eta} }  \log  M_{\text{dy},\boldsymbol \chi=0\mid \boldsymbol \nu  }(\tau)  ,
\label{eq:dynCumulantFirst}  \nonumber \\
\left[	 \boldsymbol  \kappa_{\text{dy},2, \boldsymbol \nu}   \right]_{\eta_1,\eta_2}   &=&  - \partial_{ \chi_{\eta_1} }\partial_{\chi_{\eta_2} }     \log  M_{\text{dy},\boldsymbol \chi=0\mid \boldsymbol \nu  }(\tau)   , 
\label{eq:dynCumulantSecond}
\end{eqnarray}
will be denoted as the dynamical cumulants in the following.

Evaluating  the Gaussian integral in Eq.~\eqref{eq:averagedMomentGeneratingFunction} using Eqs.~\eqref{eq:gaussianProbabilityDistribution}  and \eqref{es:conditioneMomentGenFunctionExpanded}, we obtain
\begin{multline}
	\log \overline M_{\boldsymbol \chi}(z)  = -i\boldsymbol \kappa_{1, \overline{\boldsymbol \nu}_z }   \cdot \boldsymbol \chi - \frac{1}{2} \boldsymbol  \chi  \boldsymbol \kappa_{2, \overline{\boldsymbol \nu}_z } \boldsymbol \chi   \\
	-  \frac{1}{2} \sum_{\eta_1,\eta_2 =\text{1,2} }  \left( \partial_{\overline \nu_{\eta_1} }  \boldsymbol \kappa_{1,  \overline{\boldsymbol \nu}_{z}}       \cdot \boldsymbol \chi \right)    \left[ \boldsymbol \Sigma_z^{2} \right]_{\eta_1\eta_2} \left( \partial_{\overline \nu_{\eta_2} } \boldsymbol \kappa_{1, \overline{\boldsymbol \nu}_{z}} \cdot \boldsymbol \chi  \right)
	\\
	+ \mathcal O(\boldsymbol \chi^3)   .
	\label{eq:conditioneMomentGenFkt:evaluated0}
\end{multline}
Likewise, evaluating Eq.~\eqref{eq:averagedMomentGeneratingFunctionAlt}  in the same order of $\chi$, we obtain
\begin{eqnarray}
\log \overline M_{\boldsymbol \chi}(z)  &= &  -i  \overline{\boldsymbol \nu}_{z+dz}    \cdot \boldsymbol \chi  \nonumber \\
&-& \frac{1}{2}  \boldsymbol \chi  \left( \text{diag}\, \overline{\boldsymbol \nu}_{z+dz}   +   \boldsymbol  \Sigma_{z+dz}^{2} \right)   \boldsymbol  \chi    +\mathcal O(\boldsymbol \chi^3)   . \nonumber \\
\label{eq:conditioneMomentGenFkt:evaluated1}
\end{eqnarray}
Equating  Eqs.~\eqref{eq:conditioneMomentGenFkt:evaluated0} and \eqref{eq:conditioneMomentGenFkt:evaluated1}  and deriving  with respect to the counting fields, we can construct flow equations for $ \overline{ \boldsymbol \nu}_{z} $ and  $ \boldsymbol \Sigma_{z}$. To this end, we introduce the position increment $dz$ in such that $\rho_{\text{A} } \mathcal A dz =1$ (atomic density $\rho_{\text{A} }$, laser cross section $\mathcal A$) , i.e., the length  $dz$ of a unit volume $\mathcal A dz$ which contains on average one atom.

(i) The first  derivative of  Eqs.~\eqref{eq:conditioneMomentGenFkt:evaluated0} and \eqref{eq:conditioneMomentGenFkt:evaluated1}  with respect to the counting fields yields the flow equation for the mean of the stochastic mean field
\begin{eqnarray}
\frac{d}{dz}\boldsymbol {\overline \nu}  
&=&     \boldsymbol  I_{\boldsymbol {\overline \nu}   } ,
\label{eq:tra4:momGenFunction} 
\end{eqnarray}
where the coefficient is given by
\begin{equation}
\left[ \boldsymbol  I_{\boldsymbol {\overline \nu}    } \right]_\eta  =   \rho_A  \mathcal  A  \partial_{\chi_{\eta} }  \log  M_{\text{dy},\boldsymbol \chi\mid \overline {\boldsymbol \nu}  }(\tau)   .
\label{eq:fluxEvaluation}
\end{equation}
Note that $\boldsymbol  I_{\boldsymbol {\overline \nu}  }  $ can depend on $\boldsymbol {\overline \nu}$ in a nonlinear fashion. 

(ii) The second derivative gives an equation for the covariance matrix of the stochastic mean field, and  reads
\begin{equation}
\frac{d}{dz}  \boldsymbol \Sigma^2 
=\boldsymbol  D_{\boldsymbol {\overline \nu} }   +   \boldsymbol   C_{\boldsymbol {\overline \nu} }  \boldsymbol   \Sigma^2   +\boldsymbol  \Sigma^2\boldsymbol   C_{\boldsymbol {\overline \nu}}  ,
\label{eq:varianceFlow} 
\end{equation}
where
\begin{eqnarray}
\left[	 \boldsymbol  D_{\boldsymbol {\overline \nu} }   \right]_{\eta_1,\eta_2}   &=&   \rho_{\text{A} }  \mathcal  A \;  \partial_{ \chi_{\eta_1} }\partial_{\chi_{\eta_2} }     \log  M_{\text{dy},\boldsymbol \chi\mid \boldsymbol \nu  }(\tau)   - \text{diag}( \boldsymbol  I_{\boldsymbol {\overline \nu}   } )  , \nonumber \\ 
\label{eq:diffusionMatrix}\\
\left[	 \boldsymbol  C_{\boldsymbol {\overline \nu} }  \right]_{\eta_1,\eta_2}   &=&  \rho_{\text{A} }  \mathcal  A\;   \partial_{\overline \nu_{\eta_1} }\partial_{\chi_{\eta_2} }  \log  M_{\text{dy},\boldsymbol \chi\mid \boldsymbol \nu  }(\tau)  ,
\label{phaseSpaceMatricx}
\end{eqnarray}
will be denoted as the diffusion matrix and the phase-space matrix in the following.  Note that we have neglected terms proportional to $ \partial_{\overline \nu_{\eta_1} } \boldsymbol \kappa_{1, \overline{\boldsymbol \nu}_{z}}  \partial_{\overline \nu_{\eta_2} } \boldsymbol \kappa_{1, \overline{\boldsymbol \nu}_{z}} \propto 1/ ( \overline \nu_{\eta_1} \overline \nu_{\eta_2})$ as they are suppressed for large mean fields $\overline \nu_{\eta_1},\overline \nu_{\eta_2}\gg 1$. 

After integration of the flow equations, to obtain the photon-number statistics  from the stochastic mean field statistics, we  evaluate
\begin{eqnarray}
\overline {\boldsymbol n}_{z} &=& \overline { \boldsymbol\nu }_{z},\nonumber \\
\boldsymbol \Sigma_{\text{tot},z}^2  &=& \text{diag}\,\overline { \boldsymbol\nu}_z +   \boldsymbol \Sigma_z^2    ,
\end{eqnarray}
according to the moment-generating function of the photonic distribution in Eq.~\eqref{eq:conditioneMomentGenFkt:evaluated1}. While the mean photon number equals the mean of the stochastic mean field, the photon covariance matrix  is a sum of the covariance matrix of the stochastic mean field and the photon-shot noise represented by $\overline { \boldsymbol\nu}_z $. Noteworthy, the flow equations for the mean  in Eq.~\eqref{eq:tra4:momGenFunction} and the variance in Eq.~\eqref{eq:varianceFlow}   are formally equivalent to the one in the recently developed Photon-resolved Floquet theory, which is an ab-initio semiclassical approach to determine the photon number statistics~\cite{Engelhardt2025}, yet, which does not distinguishing between photon number and stochastic mean field. As Eq.~\eqref{eq:tra4:momGenFunction}  is equivalent to the  Maxwell-Bloch equation (which can be shown along the same lines as in~\cite{Engelhardt2025}), the covariance flow equation~\eqref{eq:varianceFlow}  can be considered as an generalization of this celebrated framework.

\subsection{Statistics of polarization measurements}

\label{sec:polarizationMeasurements}
In this appendix, we provide practical guidelines how to evaluate the flow equations in Eqs.~\eqref{eq:tra4:momGenFunction}  and \eqref{eq:varianceFlow} for a polarization measurement. The  measurement statistics of the polarization rotation in a magnetometer  can be described using two photonic modes representing the two polarization directions of the linearly polarized light [i.e., the $\text{1}$ and $\text{2}$ directions in Fig.~\ref{figOverview}(a)]. We denote the photon-number operators by $\hat n_{\text{1}}$ and $\hat n_{\text{2}}$. The related stochastic mean fields are $\nu_{\text{1}}$ and $\nu_{\text{2}}$.  Since the quantum trajectory approach developed in this work  is closely related to the semiclassical approach of the Photon-resolved Floquet theory, we refer to Ref.~\cite{Engelhardt2025} for more detailed explanations, and explain here merely the key steps.

It is convenient to describe the mean photon numbers in the form
\begin{eqnarray}
\left(
\begin{array}{c}
\overline n_{\text{1}} \\
\overline n_{\text{2} }
\end{array}
\right) 
=
\overline n_{+}	\left(
\begin{array}{c}
\cos^2(\frac{\pi}{4}+ \overline\theta) \\
\sin^2(\frac{\pi}{4}+ \overline \theta)
\end{array}
\right) ,
\label{eq:meanPhotonNumbersAFOangle}
\end{eqnarray}
where $\overline n_{+}	  = \overline n_{\text{1}} + \overline n_{\text{2}} $ is the mean total photon number, and $\overline\theta$ is the mean rotation angle of the polarized light. These two quantities can be obtained by integration of  the flow equations in a  measurement basis rotated by $\overline \theta$: Denoting the corresponding photon number operators by ${\overline n}_{\text{rot}, \text{1} }$ and ${\overline n}_{\text{rot}, \text{2} }$, the rotated measurement basis fulfills $ {\overline n}_{\text{rot}, \text{1} }={\overline n}_{\text{rot}, \text{2} }$ for all $z$. In this basis, we obtain 
\begin{eqnarray}
\frac{d\overline \theta }{dz}  
&=& \frac{\partial_z {\overline n}_{\text{rot}, \text{1} } -\partial_z {\overline n}_{\text{rot}, \text{2} }  } {2\overline n_{+ }} , \nonumber \\
\frac{d{\overline n}_{+ } }{dz}  &=&\partial_z  {\overline n}_{\text{rot},\text{1} } + \partial_z {\overline  n}_{\text{rot}, \text{2}},
\label{eq:rotationAngelPhotonNumber}
\end{eqnarray}
where
\begin{equation}
\partial_z {\overline n}_{\text{rot}, \eta }  =\left[   \boldsymbol  I_{\boldsymbol {\overline \nu_{\text{rot}}}   }\right]_{\eta},
\end{equation}
and  $\boldsymbol  I_{\boldsymbol {\overline n_{\text{rot}}}   }$ is the coefficient in Eq.~\eqref{eq:fluxEvaluation}, but  evaluated in the rotated measurement basis.

Now, we  assume that
\begin{equation}
\breve I_\eta = \frac{\left[   \boldsymbol  I_{\boldsymbol {\overline \nu_{\text{rot}}}   }\right]_{\eta}   }{\overline n_{+} \rho_A \mathcal A} = \text{const.},
\label{eq:renormalizedFlux}
\end{equation}
which  is justified for off-resonant probe fields as in optical magnetometers. This  physically means that $\breve I_\alpha $ does not depend on the intensity of the probe field. In doing so, the flow equations become
\begin{eqnarray}
\frac{d\overline \theta }{dz}  
&=& \frac{\rho_A  \mathcal  A}{2} \left(  \breve  I_{\text{1}}  - \breve  I_{\text{2}} \right) , \nonumber \\
\frac{d{\overline n}_{+ } }{dz}  &=& \rho_A  \mathcal  A \left(  \breve I_{\text{1}}  +  \breve I_{\text{2}}\right)\overline n_{+ },
\end{eqnarray}
whose analytical integration yields
\begin{eqnarray}
\overline \theta(z) &=& \frac{1}{2}\rho_A  \mathcal  Az \left(  \breve  I_{\text{1}}  - \breve  I_{\text{2}} \right) ,\nonumber \\
{\overline n}_{+ }(z) &=&  \overline n_{+}^{(i)}   e^{ \left(  \breve I_{\text{1}}  +  \breve I_{\text{2}}\right) \rho_A  \mathcal  A z },
\label{eq:integratedMeans}
\end{eqnarray}
where $ \overline n_{+}^{(i)} $ denotes the total photon number before the light-matter interaction.  We note that Eq.~\eqref{eq:integratedMeans} is equivalent to Eq.~\eqref{eq:semiclassicalPolarizationRotation}  when using the integral form of the coefficients (see Sec.~\ref{sec:cumulantsIntegralForm}) and the fact that $N = \rho_A \mathcal A z $.

We are interested in the statistics of the photon-number difference
\begin{eqnarray}
\hat  n_{\text{rot},-}  &=&  \hat n_{\text{rot}, \text{1} } -{\hat n}_{\text{rot}, \text{2}} \nonumber \\
&=& \boldsymbol v_{-}  \cdot\left(
\begin{array}{c}
\hat n_{\text{rot},\text{1}} \\
\hat n_{\text{rot},\text{2} }
\end{array}
\right) ,
\end{eqnarray}
with  $\boldsymbol v_{-}  = \left(1,-1 \right)$, from which we want to infer the unknown parameter $X$. Deriving the mean value of $\hat  n_{\text{rot},-} $ with respect to $X$ yields
\begin{eqnarray}
\frac{d \overline n_{\text{rot},-}}{dX}  
&=& \overline n_{\text{rot},+} \frac{d\overline \theta}{dX} +\frac{d\overline n_{\text{rot},+} }{dX} \overline \theta(z) \nonumber \\
&=&\rho_A  \mathcal  A z	\frac{e^{ \left(  \breve I_{\text{1}}  +  \breve I_{\text{2}}\right) \rho_A  \mathcal  A z }}{2} \frac{d}{dX}\left(  \breve  I_{\text{1}}  - \breve  I_{\text{2}} \right) \nonumber \\
&+& (\rho_A  \mathcal  Az)^2	\frac{e^{ \left(  \breve I_{\text{1}}  +  \breve I_{\text{2}}\right) \rho_A  \mathcal  A z }}{2}\left(  \breve  I_{\text{1}}  - \breve  I_{\text{2}} \right) \frac{d}{dX}\left(  \breve  I_{\text{1}}  + \breve  I_{\text{2}} \right).\nonumber \\
\label{eq:derivativePhotonDifference}
\end{eqnarray}
Note that we refer here to the derivative in a fixed rotated measurement basis specified by $\overline \theta$, as otherwise the derivative would exactly vanish because of the defining property $ {\overline n}_{\text{rot}, \text{1} }={\overline n}_{\text{rot}, \text{2} }$. The last term in Eq.~\eqref{eq:derivativePhotonDifference} describes  photon absorption, and is vanishingly small for off-resonant measurements in an optical magnetometer.

To estimate the SNR $\mathcal S_X^{(n_{-})}$, we must evaluate the noise of $ \hat  n_{\text{rot},-} $.  Using the close relation to the semiclassical treatment in Ref.~\cite{Engelhardt2025}, the noise can be evaluated using the following flow equation for the covariance matrix in the  measurement basis rotated by $\overline \theta$:
\begin{eqnarray}
\frac{d}{dz}  \boldsymbol \Sigma_{\text{rot}}^2  &=& \boldsymbol  D_{\boldsymbol {\overline \nu}_{\text{rot}}}   +  \tilde {\boldsymbol   C}_{\boldsymbol {\overline \nu}_{\text{rot}}}    \boldsymbol   \Sigma^2_{\text{rot}}   +\boldsymbol  \Sigma^2_{\text{rot} } \tilde { \boldsymbol  C}_{\boldsymbol {\overline \nu}_{\text{rot}}} ,  \nonumber \\
\tilde {\boldsymbol   C}_{\boldsymbol {\overline \nu}_{\text{rot}}} &=&  {\boldsymbol   C}_{\boldsymbol {\overline \nu}_{\text{rot}}} -  \frac{d\overline \theta }{dz}  \hat  {\boldsymbol \Xi} ,
\label{eq:correlationMatrixFlow-rotated}
\end{eqnarray}
where $\hat  {\boldsymbol \Xi } = \hat \sigma_{\text{z}} + i\sigma_{\text{y} } $. To proceed, we make the following assumption for the phase-space matrix and the diffusion matrix
\begin{eqnarray}
\tilde {\boldsymbol   C}_{\boldsymbol {\overline \nu}_{\text{rot}}}  &=&  \rho_A \mathcal A  \left( \breve I_{\text{1}}  +  \breve I_{\text{2}} \right) \mathbbm 1, \label{eq:simplifyingAssumptionC} \\
\breve D&=& \frac{1}{\overline n_{+}^2} \frac{1}{\rho_A \mathcal A}	\boldsymbol v_{-}^T \boldsymbol D_{\overline{ \boldsymbol \nu} } \boldsymbol v_{-} =\text{const.},
\label{eq:simplifyingAssumptions}
\end{eqnarray}
which is fulfilled for an off-resonant probe field (see also Sec.~\ref{sec:analyticalCalculations}).

The variance  of $\hat n_{\text{rot},-}$ is related to the covariance matrix in Eq.~\eqref{eq:correlationMatrixFlow-rotated} via
\begin{equation}
\Sigma_{-}^2  = \boldsymbol v_{-}^T \boldsymbol \Sigma_{\text{rot}}^2\boldsymbol v_{-}.
\end{equation}
For the assumptions in Eq.~\eqref{eq:simplifyingAssumptions}, the variance thus fulfills the following differential equation
\begin{eqnarray}
\frac{d}{dz}   \Sigma_{-}^2  &=& \rho_A \mathcal A\breve  D\overline n_{+}^2  + 2\rho_A \mathcal A  \tilde  C \Sigma_{-}^2 ,
\label{eq:varianceFlowEquation}
\end{eqnarray}
where $\overline n_{+} (z)$ is given in Eq.~\eqref{eq:integratedMeans} and $\tilde  C  = \breve I_{\text{1}}  +  \breve I_{\text{2}}  $.
Analytical integration of  Eq.~\eqref{eq:varianceFlowEquation}  yields
\begin{eqnarray}
\Sigma_-^2(z)  
&=&  \rho_A \mathcal A z \breve D \,\overline {n}_+^2(0) .
\end{eqnarray}
We recall that this represents the fluctuations of the stoachastic mean field. To obtain the photon-number noise, we have to add the photon-shot noise, such that we obtain
\begin{eqnarray}
\Sigma_{\text{tot},-}^2(z)  
&=& {\overline n}_{+ }(z)   +   \rho_A \mathcal A z \breve D \,\overline {n}_+^2(0) ,
\label{eq:noiseAnalystical}
\end{eqnarray}
which is equivalent to Eq.~\eqref{eq:semiclassicalVariance}  after expressing $ \breve D$ as an integral as explained in Sec.~\ref{sec:cumulantsIntegralForm}.

\subsection{Master equation for the four-level system}

\label{sec:fourLevelSystemSemiclassical}

Following the procedure in Sec.~\ref{sec:masterEquation}, we  construct the generalized master equation in Eq.~\eqref{eq:generalizedMasterEquation}. In doing so, the semiclassical Hamiltonian corresponding to the microscopic Hamiltonian in Eq.~\eqref{eq:hamiltonian} is given by
\begin{eqnarray}
\hat {\mathcal H}_{m,\boldsymbol \chi}(t)    &=&  \hat H_{\text{M},m} \nonumber \\
&+&   \sum_{a=\pm}  \frac{\Omega_{a,0}}{2}  \left| e_{-a} \right> \left<g_a\right|  e^{-i\omega_{\text{p}} t}    \nonumber \\ 
&+ &   \sum_{a=\pm}  \frac{\Omega_{a,\boldsymbol  \chi}^* }{2}   \left| g_{a} \right> \left<e_{-a}\right|  e^{i\omega_{\text{p}} t}   ,
\label{eq:semiclassicalHamiltonian}
\end{eqnarray}
where 
\begin{equation}
\Omega_{a,\boldsymbol \chi}  = \Omega_{ \text{1}}e^{  i\chi_{\text{1}} }   + i a  \Omega_{ \text{2}}e^{  i\chi_{\text{y} }} 
\label{eq:countingFieldDependentRabiFrequency}
\end{equation}
with $ \Omega_{\eta} = 2 G_{\nu}  \alpha_{\eta} $ denoting the Rabi frequencies. Moreover, we assume a real-valued photonic amplitude $\alpha_\eta = \sqrt{\nu_\eta dt/\tau } $, where $\nu_\eta=\overline n_\eta$ is the  stochastic mean field. For this choice, the coherent photonic field is indeed linearly polarized.  The  counting fields $ \chi_{\text{1}},\chi_{\text{2}}$ are responsible for tracking the time-integrated measurement statistics of the photon numbers in the x and y polarization directions.

The counting-field-dependent dissipative Liouvillian is given by
\begin{eqnarray}
\mathcal L_{\boldsymbol \chi}^{\text{D}}\rho =   \gamma_{\text{z}}  \sum_{\eta=\text{1,2}} D_{\chi_\eta }\left[ \sum_{a=\pm 1} e^{i a(\eta-1) \frac{\pi}{2} }  \left| g_{m,a}\right> \left< e_{m,-a}  \right|     \right] \rho,\nonumber \\
\label{eq:liobillianGammz}
\end{eqnarray}
where $\gamma_{\text{z}} =   \Omega_{\eta}^2/ 4\dot {\overline \nu}_\eta $ according to Eq.~\eqref{eq:masterEquationConsistency}.  It connects states with different z-projection quantum numbers $a$. 
The generalized dissipator is defined in Eq.~\eqref{eq:generalizedDissipator}.

The remaining dissipation terms are given by
\begin{eqnarray}
\mathcal L^{\text{D}}\rho  &=&   2\gamma_{\text{D}} \sum_{a=\pm} D[ \left| g_{a} \right> \left<e_{-a}\right|   ] \rho \nonumber \\
&+&    \gamma_{\text{D}} \sum_{a=\pm1} D[ \left| g_{a} \right> \left<e_{a}\right|   ] \rho \nonumber\\
&+& \gamma_{\text{P} }  D\left[\frac{1}{2} \left(\left| g_{1}\right>  + \left| g_{-1} \right> \right) \left( \left< g_{1} \right|-\left< g_{-1} \right| \right)  \right]\rho,\nonumber \\
\end{eqnarray}
where $\gamma_{\text{D}}$ is the spontaneous decay rate in a particular spatial direction ($\text{x}$, $\text{y}$, or $\text{z}$). Moreover, we have phenomenologically added a pumping term $\propto \gamma_{\text{P} }$, which polarizes the atoms in the positive x direction.

An optical  magnetometer infers the strength of an external magnetic field by measuring the polarization rotation of linearly-polarized light as explained in Sec.~\ref{sec:polarizationMeasurements}.
To evaluate the signal in Eq.~\eqref{eq:derivativePhotonDifference} and the noise in Eq.~\eqref{eq:noiseAnalystical}, we need the coefficients $\breve I_{\eta}$ and $\breve D$. The dynamical moment-generating function is given by
\begin{eqnarray}
\log\,  M_{\text{dy},\boldsymbol \chi\mid \boldsymbol \nu  }(\tau)&=& \ln \text{tr} \left[  \rho_{\boldsymbol \chi} (\tau)\right] \nonumber \\ 
&=&  \lambda_{0,\boldsymbol \chi} \tau +\mathcal O (\tau^0),
\label{eq:momentGenFunctionLongTime}
\end{eqnarray}
where $\rho_{\boldsymbol \chi}$ is the reduced density matrix in Eq.~\eqref{eq:generalizedMasterEquation}. In the second line, we have used the asymptotic behavior of the dynamical moment-generating function, where $\lambda_{0,\boldsymbol \chi}$ is the dominating eigenvalue of the Liouvillian in Eq.~\eqref{eq:generalizedMasterEquation}. Using  Eq.~\eqref{eq:fluxEvaluation} and \eqref{eq:diffusionMatrix}, we can evaluate the desired coefficients $\breve I_{\eta}$ in Eq.~\eqref{eq:renormalizedFlux} and $\breve D$ in Eq.~\eqref{eq:simplifyingAssumptions}. In the numerical calculations,  we also verified that Eq.~\eqref{eq:simplifyingAssumptionC}  is fulfilled. In doing so, we have calculated the results in Figs.~\ref{figOverview} and Fig.~\ref{figSemiclassicalModel} for the semiclassical four-level system.

Of note, as we are interested in $\breve I_\pm = \breve I_{\text{1} } \pm \breve I_{\text{2} }$, we can replace $\chi_{\text{1} } = \chi_+ +\chi_-$ and $\chi_{\text{2}} = \chi_+ - \chi_-$ in the Liouvillian Eq.~\eqref{eq:generalizedMasterEquation}, and derive the generating function with respect to $\chi_\pm$ to obtain $\breve I_{\pm}$. This can be easily shown via the chain rule of derivation. Likewise,  $\breve D$ is given by the second derivative with respect to $\chi_-$.

\subsection{Effective two-level system}

\label{eq:effecitveTwoLevelSystem}

To enable an analytical investigation, we derive here an effective two-level system, which inherits the essential physical properties of the four-level system in Eq.~\eqref{eq:semiclassicalHamiltonian}. To this end, we first construct the equations of motions of the density matrix elements
\begin{eqnarray}
\frac{d}{dt} \rho^{\text{gg} }_{a,b} &=&  -i h_{L,a,-a} \rho^{\text{gg} }_{-a,b}  + i h_{R,b,-b}^* \rho^{\text{gg} }_{a,-b}\nonumber  \\
&-& i \frac{\Omega_{ a ,\boldsymbol \chi}^*  }{2}  \rho^{\text{eg} }_{-a,b}  + i \frac{ \Omega_{
		b ,-\boldsymbol \chi } }{2} \rho^{\text{ge} }_{a,-b} \nonumber \\
&+& 2\gamma_{\text{D}}  \rho^{\text{ee} }_{-a,-a} \delta_{a,b}  + \gamma_{\text{D}} \rho^{\text{ee} }_{a,a} \delta_{a,b} \nonumber\\
&+&  \gamma_{\text{P}} \left( \dots\right) +\gamma_{\text{z}} \left( \dots\right)   ,\nonumber\\
\frac{d}{dt} \rho^{\text{eg} }_{a,b} &=&  -i \epsilon_{\Delta } \rho^{\text{eg} }_{a,b}    + i h_{R,b,-b}^* \rho^{\text{eg} }_{a,-b} \nonumber \\
&-& i \frac{ \Omega_{ -a ,\boldsymbol 0 }     }{2}  \rho^{\text{gg} }_{-a,b}  + i \frac{  \Omega_{ b ,-\boldsymbol \chi}^*    }{2} \rho^{\text{ee} }_{a,-b}\nonumber  \\
&-&\frac{3 \gamma_{\text{D}}}{2} \rho^{\text{eg} }_{a,a} \delta_{a,b}\nonumber \\
&+&  \gamma_{\text{P}} \left( \dots\right) +\gamma_{\text{z}} \left( \dots\right)  ,\nonumber \\
\frac{d}{dt} \rho^{\text{ee} }_{a,b} 	&=& -i\frac{ \Omega_{ -a ,\boldsymbol 0 }      }{2}  \rho^{\text{ge} }_{-a,b}  + i \frac{ \Omega_{ -b ,-\boldsymbol 0}^*       }{2} \rho^{\text{eg} }_{a,-b} \nonumber \\
&-& (3\gamma_{\text{D} } + \gamma_{\text{z}} )\rho^{ee}_{a,a} \delta_{a,b} ,
\end{eqnarray}
where $h_{L,a,b}$ and $h_{R,a,b}$  are the matrix elements of the Hamiltonian~\eqref{eq:semiclassicalHamiltonian} in the ground-state manifold. We have not explicitly specified the terms $\propto \gamma_{\text{P}}$ and $\propto\gamma_{\text{z}}$ for brevity. The adiabatic approximation assumes that $d \rho^{\text{eg} }_{a,b}/dt  =0 $, such that
\begin{eqnarray}
\rho^{\text{eg} }_{a,b}  = \left( \rho^{ge}_{a,b} \right)^*
&\approx & \frac{1}{2} \frac{-i  \Omega_{ -a ,\boldsymbol 0 }    \rho^{\text{gg} }_{-a,b}}{ i \epsilon_{\Delta } + \frac{\Gamma}{2}}   +\mathcal O \left( \frac{\gamma_{\text{P}} \Omega}{\epsilon_{\Delta}^2 },\frac{ \Omega^2}{\epsilon_{\Delta}^2   }\right) \nonumber \\
\end{eqnarray}
for large detunings $\epsilon_{\Delta}$, where $\Gamma =3\gamma_{\text{D}} + \gamma_{\text{z}}$ is the total relaxation rate. In the main text, we have expressed this in terms of $\gamma = 3 \gamma_{\text{D}}$, which is the joint spontaneous decay rate in all three spatial directions. In the same spirit, we  also approximate the occupation of the excited-state manifold as
\begin{eqnarray}
\rho^{\text{ee} }_{a,a} 
&\approx &  \frac{1}{4} \frac{ \left| \Omega_{ -a ,\boldsymbol 0 } \right|^2  }{\epsilon_\Delta^2 + \frac{\Gamma^2}{4} }  \rho^{\text{gg} }_{-a,-a}  .
\end{eqnarray}
Inserting these expressions into the equations of motions of the ground states, we obtain the effective generalized master equation in the ground-state manifold
\begin{eqnarray}
\frac{d}{dt} \rho_{\boldsymbol \chi}  
&=&  -i \left[  \hat {\mathcal H}_{L  } \rho_{\boldsymbol \chi}   -  \rho_{\boldsymbol \chi}   \hat {\mathcal H}_{R }  \right]  \nonumber \\
&+&  	\gamma_{\text{P}} D\left[\frac{1}{2}\left( \hat \sigma_{z}+ i \hat \sigma_{y} \right)\right] \rho_{\boldsymbol \chi} \nonumber \\
&+&  	\frac{2\gamma_{\text{D}}\Omega^2 }{4\epsilon_\Delta^2 + \Gamma^2} \hat \sigma_+  \rho_{\boldsymbol \chi}\hat \sigma_-  \nonumber \\
&+&  	\frac{2\gamma_{\text{D}}\Omega^2 }{4\epsilon_\Delta^2 + \Gamma^2}  \hat  \sigma_- \rho_{\boldsymbol \chi} \hat \sigma_+  \nonumber \\
&+&  	\frac{\gamma_{\text{D}}\Omega^2 }{4\epsilon_\Delta^2 + \Gamma^2} \sum_a \ \hat \pi_a \rho_{\boldsymbol \chi} \hat \pi_a  \nonumber \\
&+&  \frac{\gamma_z }{4\epsilon_{\Delta}^2 }    \sum_{a_1,a_2=\pm 1 }   \Omega_{a_1,\boldsymbol  0} \Omega_{a_2,\boldsymbol  0}^*   \hat \pi_{a_1 }\rho_{\boldsymbol \chi} \hat \pi_{a_2 } f_{a_1,a_2}(\boldsymbol \chi)\nonumber \\
&-&	  \frac{1}{2}\frac{\Gamma }{4\epsilon_\Delta^2 + \Gamma^2 }    \sum_a \Omega_{ a ,\boldsymbol 0}  \Omega_{ a ,\boldsymbol \chi}^*    \hat \pi_a  \rho_{\boldsymbol \chi}\nonumber \\
&-&	\frac{1}{2} \frac{\Gamma  }{4\epsilon_\Delta^2 + \Gamma^2}    \sum_a \Omega_{ a ,\boldsymbol 0}^*  \Omega_{ a ,-\boldsymbol \chi}   \rho_{\boldsymbol \chi}\hat \pi_a  , 
\label{eq:genMasterEquationTwoLevelSys}
\end{eqnarray}
where 
\begin{eqnarray}
\hat {\mathcal H}_{L }  &=& - \frac{1}{2} \boldsymbol h_L \cdot \hat {\boldsymbol \sigma},  \nonumber \\
\hat {\mathcal H}_{R }   &=&    - \frac{1}{2}  \boldsymbol h_R \cdot \hat {\boldsymbol \sigma} .
\end{eqnarray}
Thereby, we have assumed a symmetric configuration  $\Omega_1 = \Omega_2 = \Omega/\sqrt{2}$.
Moreover, $\hat {\boldsymbol \sigma}  =\left(\hat \sigma_{\text{x}} ,\hat \sigma_{\text{y}} ,\hat \sigma_{\text{z}} \right)$ is a vector of Pauli matrices, $\hat \sigma_{\pm} = \sigma_{\text{x}} \pm i\sigma_{\text{y}}$, $\hat \pi_a = 1/2 +a \hat \sigma_{\text{z}}$ are projectors on the ground state $a=\pm 1$, and the entries of the vectors $\boldsymbol h_L$ and $\boldsymbol h_R $ are given by
\begin{eqnarray}
h_{L,\text{x}}  &=& h_{R,\text{x}}  = \mu B_{\text{x}} ,\nonumber \\
h_{L,\text{y}}  &=& h_{R,\text{y}}  = \mu B_{\text{y} } ,\nonumber \\
h_{L,\text{z}} 
&=&   \mu (B_{\text{z}} +\delta)   + \mathcal C_{\boldsymbol \chi}  -\mathcal C_{-\boldsymbol \chi}^* , \nonumber \\
h_{R,\text{z}} 
&=&   \mu (B_{\text{z}}-\delta )  -  \mathcal C_{\boldsymbol \chi} + \mathcal C_{-\boldsymbol \chi}^* , 
\end{eqnarray}
where
\begin{equation}
\mathcal C_{\boldsymbol \chi}  =  \frac{\epsilon_\Delta }{ 4\epsilon_\Delta^2 + \Gamma^2   }  \left(\Omega_{\text{1}}  + i \Omega_{\text{2} }   \right) \left(\Omega_{\text{1}}  e^{-i\chi_{\text{1}}  } - i \Omega_{\text{2}} e^{-i\chi_{\text{2}}  }   \right) 
\label{eq:effectivePhotonTransfer}
\end{equation}
tracks the influence of the excited states. Note that we have assumed real valued $\Omega_{\text{1}}, \Omega_{\text{2}}$ (corresponding to linearly polarized light) to reach this expression.
Moreover, we have defined
\begin{equation}
f_{a_1,a_2} (\boldsymbol \chi) = \frac{1}{2}\sum_{\eta=  1, 2} e^{i\left(a_1 -a_2\right) (\eta-1) \frac{\pi}{2}   -i\chi_{\eta }} 
\label{eq:auxilliaryFunction}
\end{equation}
for brevity, which fulfills $  f_{a_1,a_2} (\boldsymbol 0) =\delta_{a_1,a_2}$.

We have also included the parameter $\delta$ as an auxiliary parameter to calculate the QFI according to Eq.~\eqref{eq:generlizedMasterEquationFisher}. It is easy to see that Eq.~\eqref{eq:genMasterEquationTwoLevelSys}  reduces to Eq.~\eqref{eq:masterEquation} for $\boldsymbol \chi = \boldsymbol 0 $ and $\delta =0$.

\subsection{Analytical calculations}

\label{sec:analyticalCalculations}

To efficiently calculate the first two cumulants in the long-time limit according to Eq.~\eqref{eq:momentGenFunctionLongTime}, we take advantage of an expansion method for the characteristic polynomial of the Liouvillian in Eq.~\eqref{eq:generalizedMasterEquation}~\cite{Engelhardt2024c}. The characteristic polynomial can be expanded as
\begin{eqnarray}
\mathcal P (z) &=& \sum a_j z^j .
\end{eqnarray}
We denote the derivatives of the expansion coefficients by
\begin{eqnarray}
a_j^{(\eta)} &=& \left. \frac{d}{d\chi_\eta}  a_j \right|_{\chi =0}, \nonumber \\
a_j^{(\eta\zeta)} &=& \left. \frac{d}{d\chi_\eta}  \frac{d}{d\chi_\zeta}a_j \right|_{\chi=0} ,
\end{eqnarray}
where $\eta,\zeta \in \left \lbrace \text{1},\text{2},+,-\right \rbrace$ depending on the observable of interest. Using these derivatives, one can show that the first derivative  of the dominating eigenvalue of the Liouvillian in Eq.~\eqref{eq:symptoticMomentGeneratingFunction} is given by
\begin{eqnarray}
\frac{\partial}{\partial\chi_\alpha}\lambda_{0;\boldsymbol \chi=0} 
&=&      \frac{  a_0^{(\eta)}  }{ a_1}  ,
\end{eqnarray}
while the second derivatives can be evaluated via
\begin{eqnarray}
\frac{\partial^2\lambda_{0;\boldsymbol \chi=0} }{\partial\chi_\alpha \partial\chi_\beta}
&=&       \frac{  a_0^{(\alpha \beta)}  }{ a_1}     -    \frac{  a_0^{(\alpha)}   a_1^{(\beta)}  + a_0^{(\beta)}  a_1^{(\alpha)}  }{ a_1^2}  +  \frac{ 2 a_2 a_0^{(\alpha) }  a_0^{(\beta) } }{ a_1^3}  .\nonumber \\
\end{eqnarray}

For $\gamma_{\text{z}}=\gamma=0$, it is possible to find a compact expression for the expansion coefficients of the characteristic polynomial:
\begin{eqnarray}
a_4 &=&  1,\nonumber \\
a_3  &=&  2 	\gamma_{\text{P}}, \nonumber \\
a_2  &=&  \frac{1}{2} \left( h_L^2  + h_R^2  \right) +  h_{\text{x}}^2 + \frac{5}{4} 	\gamma_{\text{P}}^{2},\nonumber \\
a_1 &=& \frac{\gamma_{\text{P}} }{2} 	\left( h_L -h_R \right)^2 +  \frac{\gamma_{\text{P}} }{2}	 h_L h_R+ 	\gamma_{\text{P}} h_{\text{x}}^{2} + \frac{	\gamma_{\text{P}}^3}{4}  , \nonumber \\
a_0 &=&  \frac{1}{16}\left( h_L^2 -h_R^2   \right)^2 + \frac{	\gamma_{\text{P}}^2}{8} \left( h_L -h_R \right)^2    +   i 	\frac {\gamma_{\text{P}} } {4} h_{\text{x}}  \left( h_L^2 -h_R^2   \right) , \nonumber \\
\end{eqnarray}
for finite $\boldsymbol \chi$ and $\delta$.

Consequently, the relevant derivatives with respect to $\boldsymbol \chi=0$ are given by
\begin{eqnarray}
a_0 &=& 0 ,\nonumber\\
a_1 & =&     \frac{	\gamma_{\text{P}} }{2} h_z^2    + 	\gamma_{\text{P}} h_{\text{x}}^{2} + \frac{	\gamma_{\text{P}}^3}{4},\nonumber \\
a_2 & =&   h_{\text{z}}^2   +   h_{\text{x}}^{2} + \frac{5}{4}	\gamma_{\text{P}}^2 , \nonumber \\
a_0^{(\eta)}& =&   i \frac{1}{2}	\gamma_{\text{P}} h_{\text{x}}   h_{\text{z}}  \,\left[ 2 h_{L}^{(\eta)} \right], \nonumber  \\
a_1^{(\eta)} & =&  0 ,\nonumber  \\
a_2^{(\eta)} & =&  0 ,\nonumber  \\
a_0^{(--)}  & =& \left( \frac{1}{2} h_{\text{z}}^2    + \frac{1}{4} 	\gamma_{\text{P}}^2 \right)   \left[2 h_{L}^{(-)}   \right]^2 .
\end{eqnarray}
Putting everything together, we find
\begin{eqnarray}
\breve I_{\eta} &=&  - i  \frac{\tau }{\overline n_{+} }  \frac{d}{d\chi_\alpha}\lambda_{0;\boldsymbol\chi=\boldsymbol 0}        \nonumber \\  
&=&  \frac{\tau }{\overline n_{+} }  \frac{2   h_{ \text{x} }   h_{\text{z} }    }{  2  h_{\text{z} }^2    + 4  h_{\text{x} }^{2} +	\gamma_{\text{P}} ^2  } \, \left[4 \text{Re} \,\mathcal C_{\boldsymbol 0}^{(\eta)}  \right]  , 
\end{eqnarray}
where $\mathcal C_{\boldsymbol 0}^{(\eta)} $ refers to the derivative of $\mathcal C_{\boldsymbol \chi} $ with respect to $\chi_\eta$, which can be utilized to obtain Eq.~\eqref{eq:semiclassicalPolarizationRotation}. The diffusion coefficient for the photon-number difference becomes
\begin{eqnarray}
\breve D  &=&-  \frac{\tau }{\overline n_{+}^2 } \frac{d^2}{d^2\chi_-}  \lambda_{0;\chi=0} \nonumber \\
&=&\frac{\tau }{\overline n_{+}^2 }   \frac{2 h_{\text{z}}^2   + 	\gamma_{\text{P}}^2     }{  2 	\gamma_{\text{P}} h_{\text{z} }^2    + 4 	\gamma_{\text{P}} h_{\text{x}}^{2} + \gamma_{\text{P}}^3  } \left[\frac{\Omega^2}{\epsilon_{\Delta}} \right]^2    \nonumber \\ \nonumber \\
&-&2\frac{\tau }{\overline n_{+}^2 }  \frac{  64 \left(  h_{\text{z}}^2   +  h_{\text{x} }^{2} + \frac{5}{4} 	\gamma_{\text{P}}^2   \right) \left( \frac{1}{2} 	\gamma_{\text{P}}  h_{\text{x} }   h_{\text{z}}   \right)^2  }{\left(  2 \gamma_{\text{P}} h_{\text{z}}^2    + 4 	\gamma_{\text{P}}h_{\text{x}}^{2} + 	\gamma_{\text{P}}^3 \right)^3   } \left[\frac{\Omega^2}{\epsilon_{\Delta}}  \right]^2     . \nonumber \\
\label{eq:diffusionCoefficientAnalytic}
\end{eqnarray}
Note that $ \mathcal C_{0} \propto \Omega^2 \propto\overline n_{+} $ such that $\breve I_{\alpha}$ and   $\breve D $ are indeed  constants according to Eqs.~\eqref{eq:renormalizedFlux} and \eqref{eq:simplifyingAssumptions}.
The phase-space matrix takes the form
\begin{equation}
{\boldsymbol   C}_{\boldsymbol {\overline n}_{\text{rot}}}    = 
\rho_A \mathcal A \left( \breve I_{\text{x} } -\breve I_{\text{y} } \right)	\left(
\begin{array}{cc}
1   & 1 \\
- 1    & -1 
\end{array}
\right),
\end{equation}
showing that the condition Eq.~\eqref{eq:simplifyingAssumptionC} is fulfilled. Using now Eq.~\eqref{eq:noiseAnalystical}, we obtain the measurement noise in Eq.~\eqref{eq:semiclassicalVariance}.

In the same fashion, we can calculate the QFI for a single atom by deriving the dominating eigenvalue two times with respect to $\delta$, yielding
\begin{eqnarray}
\mathcal I_{B_{\text{z}}}^{(Q)}  &=&-  \tau  \frac{d^2}{d^2\delta}  \lambda_{0;\delta=0} \nonumber \\
&=&\tau   \frac{8 h_{\text{z}}^2   + 4	\gamma_{\text{P}}^2     }{  2 	\gamma_{\text{P}} h_{\text{z}}^2    + 4 	\gamma_{\text{P}} h_{\text{x} }^{2} + 	\gamma_{\text{P}}^3 }  \mu^2   \nonumber \\ \nonumber \\
&-&2 \tau \frac{  \left(  h_{\text{z}}^2   +  h_{\text{x} }^{2} + \frac{5}{4} 	\gamma_{\text{P}}^2   \right) \left( 2 	\gamma_{\text{P}}  h_{\text{x} }   h_{\text{z}}   \right)^2  }{\left(  2 	\gamma_{\text{P}} h_{\text{z}}^2    + 4 	\gamma_{\text{P}} h_{\text{x}  }^{2} + 	\gamma_{\text{P}}^3 \right)^3   }  \mu^2       , \nonumber \\
\end{eqnarray}
which is remarkably similar to the noise in Eq.~\eqref{eq:diffusionCoefficientAnalytic}.

\subsection{Integral expression of the cumulants}

\label{sec:cumulantsIntegralForm}

In Appendix~\ref{sec:masterEquation}, we explained how to calculate the cumulants of the photonic probability distribution using the generalized master equation. For completeness, we show here how the corresponding spectroscopic coefficients in Eqs.~\eqref{eq:renormalizedFlux}, \eqref{eq:simplifyingAssumptions} can be expressed in an  integral form  as given in the main text.

To this end, we distribute the terms in the generalized master equation in Eq.~\eqref{eq:generalizedMasterEquation} in the following form
\begin{eqnarray}
\frac{d}{dt}\rho_{\boldsymbol \chi}  
&=& \mathcal L_0 \rho_{\boldsymbol  \chi}  +\mathcal L_{1,\boldsymbol  \chi}\rho_{\boldsymbol  \chi} ,
\label{eq:generalziedMastereEquationDistribution}
\end{eqnarray}
where the Liouvillian $\mathcal L_0$ describes the dynamics of the density matrix for $\boldsymbol  \chi=0$, and the counting-field-dependent terms are contained in $\mathcal L_{1, \boldsymbol  \chi}$.

A formal solution of Eq.~\eqref{eq:generalziedMastereEquationDistribution} is given by
\begin{eqnarray}
\rho_{\boldsymbol  \chi} (t) &=& \mathcal T \exp \left[ \int_{0}^{t}dt_1 \tilde{\mathcal L}_{1,\boldsymbol  \chi}(t_1)\right]\rho(0) \nonumber \\
&=& \rho(0) + \int_{0}^{\tau} dt \tilde{\mathcal L}_{1,\boldsymbol  \chi} (t)\rho(0) \nonumber\\
&+& \int_{0}^{t}dt_1 \int_{0}^{t_1}dt_2 \tilde{\mathcal L}_{1,\boldsymbol  \chi} (t_1) \tilde{\mathcal L}_{1,\boldsymbol  \chi} (t_2)\rho(0) \nonumber\\
&+& \dots,
\end{eqnarray}
where $\mathcal T$ denotes the time-ordering operator, and
\begin{eqnarray}
\tilde{\mathcal L}_{1,\boldsymbol  \chi}(t)  \equiv  	\mathcal U^{-1}(t)  \mathcal L_{1,\boldsymbol  \chi} (t) 	\mathcal U(t)
\end{eqnarray}
is the counting-field-dependent Liouvillian transformed into an interaction picture defined by
\begin{equation}
\mathcal U(t) =  e^{\mathcal L_0 t }.
\end{equation}
Thus, the dynamical moment-generating function is given by
\begin{eqnarray}
M_{\text{dy},\boldsymbol  \chi}(t) &=& 1 + \int_{0}^{\tau}dt  \left<  \tilde{\mathcal L}_{1,\boldsymbol  \chi}(t) \right> \nonumber \\
&+& \int_{0}^{\tau}dt_1 \int_{0}^{t_1}dt_2  \left< \tilde{\mathcal L}_{1,\boldsymbol  \chi}(t_1) \tilde{\mathcal L}_{1,\boldsymbol  \chi}(t_2)\right> 
\nonumber\\
&+& \dots,
\label{eq:momentGenFktExpansion}
\end{eqnarray}
where the expectation values $\left< \bullet \right>$ must be evaluated with respect to $\rho_{\text{M}}(0)$. 

Using Eq.~\eqref{eq:renormalizedFlux} and Eq.~\eqref{eq:momentGenFktExpansion}, we can now find integral expressions for the coefficients: 
\begin{eqnarray}
\breve I_{\eta}  &=& -\frac{i}{\overline n_+  } \partial_{\chi_\eta }\log M_{\text{dy},\boldsymbol  \chi=0}, \nonumber \\
&=& \frac{1}{\overline n_+ } \int_{0}^{\tau}  dt  \left< \hat j_{\eta} (t)\right> , %
\label{eq:transportCoefficient}
\end{eqnarray}
where we have introduced the photon-flux operators
\begin{eqnarray}
\hat j_{\eta}   &=& -i\partial_{\chi_{\eta} }  \mathcal L_{\boldsymbol \chi=0}  , 
\end{eqnarray}
for $\eta=\text{1,2}$. Similarly, using the definitions in Eqs.~\eqref{eq:diffusionMatrix} and ~\eqref{eq:renormalizedFlux}, we find
\begin{eqnarray}
\breve D   
&=& \frac{1}{\overline n_{+}^2}  \int_{0}^{\tau}dt_1 \int_{0}^{t_1}dt_2  \left<	\Delta \hat j_-(t_1) 	\Delta \hat j_- (t_2)\right> ,
\label{eq:noiseCoefficient}
\end{eqnarray}
when deriving Eq.~\eqref{eq:momentGenFktExpansion} with respect to $\chi_-= \chi_{\text{1}} - \chi_{\text{2}} $.
Thereby,   $\Delta \hat j_{-}(t) =  \hat j_{-}(t) - \langle \hat j_{-} (t) \rangle $ with $\hat j_{-} = \hat j_{\text{1}}  - \hat j_{\text{2}} $.  Interestingly, careful comparison of Eqs.~\eqref{eq:diffusionMatrix}, \eqref{eq:semiclassicalHamiltonian}, and \eqref{eq:liobillianGammz} reveals that the first term in Eq.~\eqref{eq:momentGenFktExpansion} is exactly canceled, and only the integrated correlation function contributes to $\breve D$.

For the effective two-level system in Eq.~\eqref{eq:genMasterEquationTwoLevelSys}, the photon-flux operators become
\begin{eqnarray}
\hat {j}_{\text{1}}\rho &=& \frac{\epsilon_\Delta \Omega^2}{ 4\epsilon_\Delta^2 + \Gamma^2   }  \frac{1}{2} \left( \hat \sigma_{\text{z}} \rho + \rho \hat \sigma_{\text{z}} \right) \nonumber \\
&+&  \frac{\gamma_{\text{z}} }{4\epsilon_{\Delta}^2 }  \frac{\Omega^2}{2}   \sum_{a_1,a_2=\pm 1 }   (1+ia_1) (1-ia_2)\hat \pi_{a_1 }\rho \hat \pi_{a_2 }\frac{1}{2} \nonumber \\
&-&	  \frac{1}{2}\frac{\Gamma }{4\epsilon_\Delta^2 + \Gamma^2 }  \frac{\Omega}{2}  \sum_a (1+ia)  \hat \pi_a  \rho\nonumber \\
&-&	\frac{1}{2} \frac{\Gamma  }{4\epsilon_\Delta^2 + \Gamma^2} \frac{\Omega}{2}   \sum_a (1-ia)  \rho\hat \pi_a  , \nonumber \\
\hat {j}_{\text{2}}\rho &=& -\frac{\epsilon_\Delta \Omega^2}{ 4\epsilon_\Delta^2 + \Gamma^2   }  \frac{1}{2} \left( \hat \sigma_{\text{z}} \rho + \rho \hat \sigma_{\text{z}} \right) \nonumber \\
&+&  \frac{\gamma_{\text{z}} }{4\epsilon_{\Delta}^2 }  \frac{\Omega^2}{2}   \sum_{a_1,a_2=\pm 1 }   (1+ia_1) (1-ia_2)\hat \pi_{a_1 }\rho \hat \pi_{a_2 }\frac{ e^{i\left(a_1 -a_2\right) }}{2} \nonumber \\
&-&	  \frac{1}{2}\frac{\Gamma }{4\epsilon_\Delta^2 + \Gamma^2 }  \frac{\Omega}{2}  \sum_a (1+ia)  ia \hat \pi_a  \rho\nonumber \\
&-&	\frac{1}{2} \frac{\Gamma  }{4\epsilon_\Delta^2 + \Gamma^2} \frac{\Omega}{2}   \sum_a (1-ia) (-ia) \rho\hat \pi_a  , 
\end{eqnarray}
which is equivalent to the expression in Eq.~\eqref{eq:photonFluxOperatorsSC}  when neglecting $\gamma_{\text{z}}$ due to its smallness.

\section{Collective model}

\label{sec:app:collectiveModel}

In this appendix, we provide the details about the calculations in the collective model. In Appendix~~\ref{eq:generalizedMasterEquation:collective}, we derive the generalized master equation. In Appendix~\ref{sec:nonUnitaryMeanFieldTheory}, we introduce a non-unitary mean-field theory to analyze the collective master equation in the thermodynamic limit. In Appendix~\ref{app:lowOrderCumulantsCalculation}, we explain how to evaluate the non-unitary mean-field theory efficiently. In Appendix~\ref{app:benchmarking}, we carry out benchmark calculations of the non-unitary mean-field theory. In Appendix~\ref{sec:integralExpressions}, we explain how to relate the full-counting statistics applied here to the integral expressions  in the main text. In Appendix~\ref{sec:collective:exact}, we show how to calculate the QFI and the SNR analytically for a vanishing magnetic field using the Bogoliubov transformation. Finally, in Appendix~\ref{sec:experimentalLimitations}, we discuss possible experimental limitations of this model.

\subsection{Generalized master equation}

\label{eq:generalizedMasterEquation:collective}

Here, we derive a generalized collective quantum master equation  for the Hamiltonian in Eq.~\eqref{eq:hamiltonian}, which is equivalent to a quantum trajectory approach.
The derivation  proceeds in two steps. First, we carry out an adiabatic elimination of the excited states~\cite{Hammerer2010}. Second, we trace out the photonic degrees of freedom to obtain a master equation acting only on the atomic ground states. The information about the photonic modes is traced by the counting fields as in Sec.~\ref{sec:masterEquation}.

To implement the adiabatic elimination, we construct the Heisenberg equations of the coherence operators for the Hamiltonian in Eq.~\eqref{eq:hamiltonian}, which read
\begin{eqnarray}
\frac{d}{dt}\left| e_{m,-a} \right> \left< g_{m,a}\right| &=& -i\epsilon_{\Delta} \left| e_{m,-a} \right> \left< g_{m,a}\right| \nonumber \\
&-& i  \sum_{\eta= \text{1,2}} G_{a,\xi}   \hat a_{\xi}^\dagger \left| g_{m,a} \right> \left< g_{m,a}\right| \nonumber \\
&-&i \sum_{\eta= \text{1,2}}  G_{a,\xi}^* \left| e_{m,-a} \right> \left< e_{m,-a}\right|   \hat a_{\xi} . \nonumber \\
\label{eq:coherenceEOM}
\end{eqnarray}
Thereby, we have parameterized the light-matter coupling to be $G_{a,\xi} = G_{\xi} e^{i a(\eta-1) \frac{\pi}{2} } $, where we recall that $\xi = (\eta ,t)$. Assuming that the excited state is unoccupied, and the left-hand side in Eq.~\eqref{eq:coherenceEOM} is zero, we find
\begin{equation}
\left| e_{m,-a} \right> \left< g_{m,a}\right| =  - \sum_{\eta= \text{1,2}} \frac{G_{a,\xi}^* }{\epsilon}  \hat a_{\xi}^\dagger \left| g_{m,a} \right> \left< g_{m,a}\right| .  
\label{eq:coherenceAdiabatic}
\end{equation}
Using this result, we can also represent the residual excited-state occupation as
\begin{equation}
\left| e_{m,a} \right> \left< e_{m,a}\right| =  \sum_{\eta_1,\eta_2} \frac{G_{a,\xi_1}^*  G_{a,\xi_2} }{\epsilon^2} \left| g_{m,-a} \right> \left< g_{m,-a}\right|  \hat a_{\xi_1}^\dagger\hat a_{\xi_2}  .
\label{eq:occupationAdiabatic}
\end{equation}
Inserting Eqs.~\eqref{eq:coherenceAdiabatic} and \eqref{eq:occupationAdiabatic} into the Hamiltonian Eq.~\eqref{eq:hamiltonian} in the Schr\"odinger picture, we obtain,
\begin{eqnarray}
\hat H  
&=& -  \mu \boldsymbol B \cdot \hat {\boldsymbol S} - \sum_{\eta_1,\eta_2} \sum_{a}   \frac{G_{a,\xi_1}^*  G_{a,\xi_2}  }{\epsilon_\Delta }  \hat P_a  \hat a_{\xi_1}^\dagger\hat a_{\xi_2}  ,
\end{eqnarray}
where  we introduced $\hat P_a \equiv  \sum_m   \left| g_{m,a} \right> \left< g_{m,a}\right|   = N/2 +a \hat S_{\text{z}}$ as the operator counting the  occupation in state $a$. Moreover, the collective angular-moment operators are defined as $\hat {\boldsymbol S}_\eta = \sum_{m} \hat \sigma_{m,\eta}/2$ for $\eta =\text{x,y,z}$.  We note that this effective Hamiltonian  neglects spontaneous photon emission due to the adiabatic elimination.

Starting from this effective Hamiltonian, we can now trace out the photonic degrees of freedom using the same procedure as in Sec.~\ref{sec:masterEquation}. In doing so, we obtain the generalized master equation
\begin{widetext}
	\begin{eqnarray}
	\frac{d}{dt} \rho_{\boldsymbol \chi}   &=&  i \left[  \mu \boldsymbol B \cdot \hat{\boldsymbol S},  \rho_{\boldsymbol \chi}  \right] -  i \sum_{\eta_1,\eta_2 } \sum_{a } \left(e^{-i\chi_{\eta_1}}\frac{ \Omega_{a,\eta_1}^* \Omega_{a,\eta_2}  }{4\epsilon_{\Delta} }   \hat P_{a } \rho_{\boldsymbol \chi}
	- e^{-i\chi_{\eta_1}} \frac{ \Omega_{a,\eta_1} \Omega_{a,\eta_2}^*  }{4\epsilon_{\Delta}}   \rho_{\boldsymbol \chi}\hat P_{a } \right)   \nonumber \\
	&+&   \sum_{\eta_1,\eta_2, \eta_3 } \sum_{a_1,a_2 }  \left( e^{-i\chi_{\eta_1}}\frac{G_{a_1,\eta_1}^* G_{a_2,\eta_1} \Omega_{a_1,\eta_2} \Omega_{a_2,\eta_3}^*  }{4 \epsilon_{\Delta}^2}   \hat P_{a_1 } \rho_{\boldsymbol \chi} \hat P_{a_2 }  \right. \nonumber \\
	&-& \left.  \frac{1}{2} e^{-i\chi_{\eta_2}}\frac{G_{a_1,\eta_1} G_{a_2,\eta_1}^* \Omega_{a_1,\eta_2}^*  \Omega_{a_2,\eta_3} }{4 \epsilon_{\Delta}^2}   \hat P_{a_1 } \hat P_{a_2 } \rho_{\boldsymbol \chi} 
	-  \frac{1}{2} e^{-i\chi_{\eta_3}}\frac{G_{a_1,\eta_1} G_{a_2,\eta_1}^* \Omega_{a_1,\eta_2}^* \Omega_{a_2,\eta_3}  }{4 \epsilon_{\Delta}^2}   \rho_{\boldsymbol \chi} \hat P_{a_1 } \hat P_{a_2 }  \right)dt \nonumber \\
	\label{eq:cal:Liouv}
	\end{eqnarray}
\end{widetext}
with  the Rabi frequencies $\Omega_{a,\eta}   = 2G_{a,\eta }\alpha_{\eta} $. Thereby, we have assumed  time-integrated intensities which is equivalent to simply replacing $\xi \rightarrow \eta $ [see Eq.~\eqref{eq:timeInvariantCountingField}]. The dissipation terms in the third, fourth, and fifth lines can be simplified using the relations
\begin{eqnarray}
\sum_{\eta=\text{1,2} }G_{a_1,\eta}^* G_{a_2,\eta}  &=& 2 \frac{\gamma_{\text{z}}}{dt} \delta_{a_1,a_2}, \nonumber \\
\sum_{\eta=\text{1,2} }\Omega_{a_1,\eta}^* \Omega_{a_2,\eta}  &=& \Omega^2 \delta_{a_1,a_2}.
\end{eqnarray} 
Moreover, we use the counting-field-dependent Rabi frequencies in Eq.~\eqref{eq:countingFieldDependentRabiFrequency} such that
\begin{eqnarray}
\frac{d}{dt} \rho_{\boldsymbol \chi}   &=&  -i \left[  \hat {\mathcal H}_{L} \rho_{\boldsymbol \chi} -   \rho_{\boldsymbol \chi} \hat {\mathcal H}_{R }  \right]  \nonumber \\
&+&    \gamma_{\text{P}}  D [ \hat S_{\text{z}} +i   \hat S_{\text{y}}   ] \rho_{\boldsymbol \chi}  \nonumber \\
&+&  \frac{\gamma_{\text{z} } }{2\epsilon_{\Delta}^2 }    \sum_{a_1,a_2=\pm 1 }   \Omega_{a_1,\boldsymbol  0} \Omega_{a_2,\boldsymbol  0}^*   \hat P_{a_1 }\rho_{\boldsymbol \chi} \hat P_{a_2 } f_{a_1,a_2} (\boldsymbol \chi)\nonumber \\
&-&  \frac{1}{2} \frac{\gamma_{\text{z} } }{2\epsilon_{\Delta}^2 }  \sum_{a =\pm 1 } \Omega_{ a ,\boldsymbol 0}  \Omega_{ a ,\boldsymbol \chi}^* \hat P_{a } \hat P_{a } \rho_{\boldsymbol \chi}\nonumber \\
&-&  \frac{1}{2} \frac{\gamma_{\text{z} } }{2\epsilon_{\Delta}^2 }    \sum_{a=\pm 1 }  \Omega_{ a ,-\boldsymbol \chi} \Omega_{ a ,\boldsymbol 0}^*   \rho_{\boldsymbol \chi} \hat P_{a } \hat P_{a } , 
\label{eq:collectiveLiouvillian}
\end{eqnarray}
where
\begin{eqnarray}
\hat {\mathcal H}_{L}  &=& - h_{\text{x}}\hat S_{\text{x}} -   \left( h_{\text{z}}+\mu \delta \right) \hat S_{\text{z} } 
+ \left(  \mathcal C_{\boldsymbol \chi}  -\mathcal C_{-\boldsymbol \chi}^* \right)\hat S_{\text{z}}  \nonumber \\
\label{eq:generalizedHamiltonian}
\end{eqnarray}
with $ \mathcal C_{\boldsymbol \chi}$  given in Eq.~\eqref{eq:effectivePhotonTransfer} (for $\Gamma =0$), and $f_{a_1,a_2} (\boldsymbol \chi)$ given in Eq.~\eqref{eq:auxilliaryFunction}.
Moreover, $\hat {\mathcal H}_{R}  = \left. \hat {\mathcal H}_{L}\right|_{\boldsymbol \chi \rightarrow -\boldsymbol \chi,\delta \rightarrow -\delta}  $.  
We have phenomenologically added a pump term $\propto \gamma_{\text{P}}$ to establish a close connection to the semiclassical model in Eq.~\eqref{eq:genMasterEquationTwoLevelSys}. Note that the semiclassical master equation in  Eq.~\eqref{eq:genMasterEquationTwoLevelSys} and the collective master equation in Eq.~\eqref{eq:collectiveLiouvillian} are formally equivalent for $\gamma_{\text{D}}=0$.

The appearance of the counting fields in the dissipators $\propto \gamma_{\text{z}}$ is uncommon. However, for $\Omega_{\text{x}} = \Omega_{\text{y}}$, the master equation can be simplified:  Using the definition of the moment-generating function as $M_{\boldsymbol \chi} =\text{tr} [\rho_{\boldsymbol \chi} ]$ and the fact that $\hat P_{a}  = N/2 +a \hat S_{\text{z}} $, one can show that the master equation \eqref{eq:collectiveLiouvillian} can be brought in the form
\begin{equation}
\frac{d}{dt} \rho_{\boldsymbol \chi}  =  \mathcal L_0 \rho_{\boldsymbol \chi}  +\sum_{\eta} e^{i\chi_\eta} \mathcal L_\eta \rho_{\boldsymbol \chi},
\end{equation}
where $\mathcal L_0$ contains the first two terms in Eq.~\eqref{eq:collectiveLiouvillian}, while the  $\mathcal L_\eta $ are proportional to $\gamma_{\text{z}}$, do not contain any counting fields, and $\text{tr}\left[\mathcal L_\eta \hat A \right] = 0$ for arbitrary operators $\hat A$. Because of this feature, one can show that the counting fields in the dissipator in Eq.~\eqref{eq:collectiveLiouvillian} do not contribute in the calculation of the stationary cumulants~\cite{Flindt2010,Hussein2014}. For this reason, we  set $\chi_{\eta} =0$ in the dissipator. Yet, we note that these terms can contribute in the transient dynamics.

Using $\hat P_{a}  = N/2 +a \hat S_{\text{z}} $, we thus obtain
\begin{eqnarray}
\frac{d}{dt} \rho_{\boldsymbol \chi}   &=&  -i \left[  \hat {\mathcal H}_{L} \rho_{\boldsymbol \chi} -   \rho_{\boldsymbol \chi} \hat {\mathcal H}_{R }  \right]  \nonumber \\
&+&    \gamma_{\text{P}}  D [ \hat S_{\text{z}} +i   \hat S_{\text{y}}   ] \rho_{\boldsymbol \chi} +    \tilde \gamma_{\text{z} }  D [   \hat S_{\text{z}}   ] \rho_{\boldsymbol \chi}
\label{eq:collectiveLiouvillianFinal}
\end{eqnarray}
with $ \tilde \gamma_{\text{z} } = \gamma_{\text{z} } \Omega^2/ \epsilon_{\Delta}^2 $,  which reduces to the master equation in Eq.~\eqref{eq:collectiveMasterEquation} for $\chi_{\eta}=\delta =0$. Using the counting-field dependent density matrix, we obtain the moment-generating function
\begin{equation}
M_{\boldsymbol \chi }(\tau)  =  M_{\text{dy},\boldsymbol \chi }(\tau)  M_{\boldsymbol \chi}(0),
\label{eq:momentGenFkt:collective}
\end{equation}
where $M_{\text{dy},\boldsymbol \chi } (\tau ) = \text{tr} \left[ \rho_{\boldsymbol \chi}  (\tau) \right]$ and $M_{\boldsymbol \chi}(0)$ given in Eq.~\eqref{eq:initialMomentGenFunction} for $\xi \rightarrow \eta $.

\subsection{Non-unitary mean-field theory}

\label{sec:nonUnitaryMeanFieldTheory}

As the ensemble consists of a macroscopic number of atoms, we carry out a mean-field treatment to analyze the measurement statistics. As a first step, we apply the   Holstein-Primakoff transformation
\begin{eqnarray}
\hat S_{\text{x}} &=& \hat a^\dagger  \hat a - \frac{N}{2} ,\nonumber  \\
\hat S_{\text{y}}  &=& \frac{1}{2i}\left( \sqrt{N -  \hat a^\dagger  \hat a  }\hat a  - \hat a^\dagger  \sqrt{N -  \hat a^\dagger  \hat a  }      \right),\nonumber  \\
\hat S_{z} &=& \frac{1}{2}\left( \sqrt{N -  \hat a^\dagger  \hat a  }\hat a  +\hat a^\dagger  \sqrt{N -  \hat a^\dagger  \hat a  }      \right),
\label{eq:bogolioubov}
\end{eqnarray}
where $\hat a$ is a bosonic operator, and $N$ denotes the total number of atoms in the ensemble.   In terms of the Holstein-Primakoff boson, the generalized master equation reads
\begin{widetext}
	\begin{eqnarray}
	\frac{d}{dt} \rho_{\boldsymbol \chi}   &=&  -i \left[  -h_{\text{x}} \left(\hat a^\dagger  \hat a - \frac{N}{2}     \right) -  \left(  \ h_{\text{z}}+\mu \delta+   \mathcal C_{\boldsymbol \chi}  -\mathcal C_{-\boldsymbol \chi}^*     \right) \frac{1}{2}\left(\hat a^\dagger  \sqrt{N -  \hat a^\dagger  \hat a  }  +   \sqrt{N -  \hat a^\dagger  \hat a  }  \hat a      \right) 
	\right]  \rho_{\boldsymbol \chi} \nonumber \\
	&+&     i \rho_{\boldsymbol \chi } \left[ -  h_{\text{x}} \left(\hat a^\dagger  \hat a - \frac{N}{2}  \right) -  \left(   h_{\text{z}}-\mu \delta-  \mathcal C_{\boldsymbol \chi}  + \mathcal C_{-\boldsymbol \chi}^*    \right)\frac{1}{2} \left(\hat a^\dagger  \sqrt{N -  \hat a^\dagger  \hat a  }  +   \sqrt{N -  \hat a^\dagger  \hat a  }  \hat a      \right) 
	\right]   \nonumber   \\
	&+&    \gamma_{\text{P}} \sqrt{N -  \hat a^\dagger  \hat a  } \hat a  \rho_{\boldsymbol \chi} \hat a^\dagger  \sqrt{N -  \hat a^\dagger  \hat a  }  -\frac{\gamma_{\text{P}} }{2}  \left( N-1 - \hat a^\dagger   \hat a \right)\hat a^\dagger   \hat a \rho_{\boldsymbol \chi} -\frac{\gamma_{\text{P} } }{2} \rho_{\boldsymbol \chi}   \left( N-1 - \hat a^\dagger   \hat a \right)\hat a^\dagger   \hat a\nonumber \\
	%
	%
	&+&	  \tilde \gamma_{\text{z} } D   \left[\frac{1}{2}\left( \sqrt{N -  \hat a^\dagger  \hat a  }\hat a  +\hat a^\dagger  \sqrt{N -  \hat a^\dagger  \hat a  }      \right)  \right] \rho_{\boldsymbol \chi} .
	\label{eq:masterEquation:HolsteinPrimakoff}
	\end{eqnarray}
\end{widetext}

In contrast to the common mean-field theory in unitary systems, where one introduces a single mean field via a shift operation $\hat a \rightarrow \hat a+\alpha$ with a complex valued $\alpha$, here we introduce four distinct complex mean fields $(\alpha_{f} ,\alpha_{f*} ,  \alpha_{b} ,\alpha_{b*} ) = \boldsymbol \alpha$, which we arrange as a vector for brevity~\cite{Li2025}. These complex mean fields define the following non-unitary shift operations
\begin{eqnarray}
\hat W_{f} &=& \exp\left[  \sqrt{N}\left( \alpha_{f*}  \hat a - \alpha_{f}\hat a^\dagger\right) \right], \nonumber \\
\hat W_{b} &=& \exp\left[  \sqrt{N} \left( \alpha_{b*}  \hat a - \alpha_{b}\hat a^\dagger \right) \right].
\end{eqnarray}
The labels $f$ and $b$  indicate that  these operators are supposed to operate on the \textit{front} or \textit{back} side of the  density matrix. Under these operations, the Holstein-Primakoff boson transforms as
\begin{eqnarray}
\hat W_{f} \hat a \hat W_{f}^{-1}  &=&\hat a + \sqrt{N} \alpha_{f} , \nonumber \\
\hat W_{f} \hat a^\dagger \hat W_{f}^{-1} &=&\hat a^\dagger +  \sqrt{N} \alpha_{f*}, \nonumber \\
\hat W_{b} \hat a \hat W_{b}^{-1}  &=& \hat a + \sqrt{N} \alpha_{b} , \nonumber \\
\hat W_{b} \hat a^\dagger \hat W_{b}^{-1}  &=& \hat a^\dagger +  \sqrt{N} \alpha_{b*} .
\end{eqnarray}
Using these shift operators, we define the non-unitary superoperator $\mathcal W$ via the action on the density matrix
\begin{equation}
\mathcal W \left(\rho \right) = \hat W_{f}     \rho   \hat W_{b}^{-1}  .\\
\end{equation}
Accordingly, the inverse transformation is  given by
\begin{equation}
\mathcal W^{-1} \left(\rho \right) = \hat W_{f}^{-1}     \rho   \hat W_{b}  .\\
\end{equation}
For instance, we have
\begin{eqnarray}
\mathcal W \left(\hat a \rho \hat a^\dagger  \right) &= & \hat W_{f}  \hat a   \rho  \hat a \hat W_{b}^{-1} \nonumber \\
&= &  \hat W_{f} \hat a \hat W_{f}^{-1}  \hat W_{f}  \rho  \hat W_{b}^{-1}  \hat W_{b} \hat a \hat W_{b}^{-1} \nonumber \\
&= & \left(\hat a + \sqrt{N} \alpha_{f } \right) \tilde \rho  \left(\hat a^\dagger + \sqrt{N} \alpha_{b*}  \right)\nonumber \\
&= & N\alpha_{f } \alpha_{b*}     \tilde  \rho   +   \sqrt{N}  \left( \alpha_{b*}   \hat a   \tilde \rho   +  \alpha_{f } \tilde \rho     \hat a^\dagger \right)   +  \hat a   \tilde \rho  \hat a ^\dagger ,\nonumber \\
\end{eqnarray}
where $\tilde \rho$ is the density matrix in the transformed frame. In the last line, we have distributed the transformed operator in orders of the atom number. In the semiclassical limit $N\rightarrow\infty$, the first term will dominate, such that only the mean-field contribution is relevant.

Applying the transformation $\mathcal W$ to the generalized master equation, we formally obtain 
\begin{eqnarray}
\frac{d}{dt}\rho_{\boldsymbol \chi} &=& \mathcal L_{\boldsymbol \chi}  \rho_{\boldsymbol \chi}\nonumber \\
&=&\mathcal W^{-1} \left[  \mathcal W \left( \mathcal L_{\boldsymbol \chi}  \rho_{\boldsymbol \chi} \right)\right]\nonumber \\
&=& N \lambda_{\boldsymbol \chi} \mathcal W^{-1} \left[ \tilde \rho_{\boldsymbol \chi}\right]  + \sqrt{N} \mathcal W^{-1} \left[ {\mathcal L_{1}} \tilde \rho_{\boldsymbol \chi}\right] +\mathcal W^{-1} \left[ {\mathcal L_{2}} \tilde \rho_{\boldsymbol \chi}\right]\nonumber \\
&=& N \lambda_{\boldsymbol \chi}  \rho_{\boldsymbol \chi} + \sqrt{N} \mathcal W^{-1} \left[ \breve{\mathcal L_{1}} \tilde \rho_{\boldsymbol \chi}\right] +\mathcal W^{-1} \left[ \breve{\mathcal L_{2}} \tilde \rho_{\boldsymbol \chi}\right].\nonumber \\
\end{eqnarray}
In the leading order $\propto N $, the Liouvillian is thus just a complex-valued number $\lambda_{\boldsymbol \chi}$. To cancel the term $\propto \sqrt{N}$, we choose the mean field such that 
\begin{eqnarray}
0 &=&	{\mathcal L_{1}} \tilde\rho_{\boldsymbol \chi} \nonumber \\
&=& l_{f}(  \boldsymbol \alpha ) \hat a\tilde\rho_{\boldsymbol \chi}  + l_{f*}( \boldsymbol \alpha ) \hat a^\dagger\tilde \rho+ l_{b}( \boldsymbol \alpha) \tilde\rho_{\boldsymbol \chi}  \hat a+ l_{b*}( \boldsymbol \alpha ) \tilde\rho_{\boldsymbol \chi} \hat a^\dagger , \nonumber \\
\label{eq:holPrim:firstOrder}
\end{eqnarray}
which gives four nonlinear equations [$l_{f}(  \boldsymbol \alpha ) = 0$ etc.] for four independent variables. These equations can be solved numerically.

In terms of the reduced density matrix, the time-evolution of the moment-generating function follows
\begin{eqnarray}
\frac{d}{dt} M_{\text{dy},\boldsymbol \chi} &=& \frac{d}{dt} \text{tr}\left[  \rho_{\boldsymbol \chi} \right] \nonumber \\ 
&=& N \lambda_{\boldsymbol \chi} M_{\text{dy},\boldsymbol \chi} + \text{tr}\left[ \mathcal W^{-1} \left( \breve{\mathcal L_{2}} \tilde\rho_{\boldsymbol \chi}  \right)\right].
\end{eqnarray}
Thus, the moment-generating function can be formally written as
\begin{equation}
M_{\text{dy},\boldsymbol \chi}(t) =  e^{N \lambda_{\boldsymbol \chi}t}\breve M_{\text{dy},\boldsymbol \chi}(t),
\end{equation}
where $\breve M_{\text{dy},\boldsymbol \chi}(t)$ contains  contributions of order $N^0$. This allows us to define an asymptotic  cumulant-generating function in the thermodynamic limit $N\rightarrow \infty$ as
\begin{eqnarray}
\overline K_{\boldsymbol \chi} &\equiv&  \lim_{N\rightarrow\infty}   \lim_{t\rightarrow\infty} \frac{1}{tN}\ln \left[M_{\text{dy},\boldsymbol \chi}(t)M_{\boldsymbol \chi }(0)  \right]  \nonumber \\
&=& \lambda_{\boldsymbol \chi} + \frac{1}{N}\sum_{\eta =1,2} \left( e^{-i\chi_\eta} -1\right) \dot{ \overline \nu}_{\eta} ,
\end{eqnarray}
according to Eq.~\eqref{eq:momentGenFkt:collective} with $\dot{ \overline \nu}_{\eta} = \overline \nu_{\eta }/\tau$ being the mean photon flux.
For the generalized master equation in Eq.~\eqref{eq:masterEquation:HolsteinPrimakoff}, we formally assume that $\gamma_{\text{P} }$ and $\tilde \gamma_{\text{z}}$ scale as $N^{-1}$ to obtain a well-defined thermodynamic limit.  Then, the asymptotic cumulant-generating function turns out to be 
\begin{widetext}
	\begin{eqnarray}
	\overline K_{\boldsymbol \chi}   &=&  -i \left[ - h_{\text{x}}   \left(\alpha_{f*}  \alpha_{f} - \frac{1}{2}     \right) -  \left(  h_{\text{z}}+\mu \delta +    \mathcal C_{\boldsymbol \chi}  -\mathcal C_{-\boldsymbol \chi}^*   \right)\frac{1}{2} \left(\alpha_{f*}   \sqrt{1 -  \alpha_{f*} \alpha_{f} }  +   \sqrt{1 -  \alpha_{f*}  \alpha_{f} }  \alpha_{f}     \right) 
	\right] \nonumber \\
	&+&     i  \left[  - h_{\text{x}} \left( \alpha_{b*}  \alpha_{b} - \frac{1}{2}  \right) -  \left(   h_{\text{z}} -\mu \delta -    \mathcal C_{\boldsymbol \chi}  + \mathcal C_{-\boldsymbol \chi}^* \right) \frac{1}{2} \left( \alpha_{b*}   \sqrt{1 -  \alpha_{b*}\alpha_{b}  }  +   \sqrt{1 -  \alpha_{b*} \alpha_{b}  }  \alpha_{b}      \right) 
	\right]   \nonumber   \\
	&+&   N  \gamma_{\text{P}}  \sqrt{1-  \alpha_{f*}  \alpha_{f}   } \alpha_{f} \alpha_{b*}   \sqrt{1 -  \alpha_{b*}  \alpha_{b}  }  -N \frac{\gamma_{\text{P} } }{2}  \left( 1- \alpha_{f*}  \alpha_{f}  \right)\alpha_{f*}  \alpha_{f}  -N\frac{\gamma_{\text{P} } }{2}  \left( 1 - \alpha_{b*}  \alpha_{b} \right)\alpha_{b*}  \alpha_{b}\nonumber \\
	%
	%
	%
	&+&  	N \frac{ \tilde \gamma_{\text{z} }}{4}   \left(\alpha_{f*}   \sqrt{1-  \alpha_{f*}   \alpha_{f}  }  +   \sqrt{1 -  \alpha_{f*}   \alpha_{f} }  \alpha_{f}     \right)  \left(\alpha_{b*}  \sqrt{1 -  \alpha_{b*}  \alpha_{b} }  +   \sqrt{1 -  \alpha_{b*}  \alpha_{b}  }  \alpha_{b}     \right)  \nonumber \\
	&-& N \frac	{ \tilde \gamma_{\text{z} }  }{8}    \left( \alpha_{f*}    \sqrt{1 -   \alpha_{f*}   \alpha_{f}   }  +   \sqrt{1 -   \alpha_{f*}   \alpha_{f}    }  \alpha_{f}        \right)  ^2 \nonumber \\
	&-&	N \frac	{ \tilde \gamma_{\text{z} } }{8}  \left( \alpha_{b*}   \sqrt{1 -   \alpha_{b*}   \alpha_{b}  }  +   \sqrt{1 -  \ \alpha_{b*}   \alpha_{b}  } \alpha_{b}     \right)^2 \nonumber \\
	&+& \frac{1}{N}\sum_{\eta=\text{1},\text{2}} \left( e^{-i\chi_\eta} -1\right) \dot{ \overline \nu}_{\eta}.
	\label{eq:masterEquation:HolsteinPrimakoffMeanfield}
	\end{eqnarray}
\end{widetext}
Notably, this procedure can be amended with arbitrary operators (e.g., $\hat S_{\text{y}}$) by including the respective Holstein-Primakoff representation.

\subsection{Efficient evaluation of low-order cumulants}

\label{app:lowOrderCumulantsCalculation}

In this appendix, we explain how to efficiently calculate low-order cumulants of the measurement statistics. Crucially, the mean-fields appearing in Eq.~\eqref{eq:masterEquation:HolsteinPrimakoff} are determined  as the roots of  $l_{f}, l_{f*}, l_{b}, l_{b*} $ in Eq.~\eqref{eq:holPrim:firstOrder}, and depend thus implicitly on the counting fields $\boldsymbol \chi$.  This fact renders the evaluation of the derivatives of $\overline K_{\boldsymbol \chi}  $ with respect to the counting fields $\chi$ numerically unstable. For this reason, a more sophisticated evaluation method is required.  For simplicity, we will focus on a single counting field in the following.

For brevity, we introduce $\alpha_1 =\alpha_{f} $,    $\alpha_2 =\alpha_{f*} $,   $\alpha_3 =\alpha_{b} $,   $\alpha_4 =\alpha_{b*} $ and likewise for the $l_{f}, l_{f*}, l_{b}, l_{b*} $. We denote the roots  of the $l_{i}$  by $\alpha_i^{(0)}$ when $\chi=0$, i.e.,
\begin{equation}
l_{i}  \left( \boldsymbol \alpha^{(0)} \right) =0.
\label{eq:meanFieldCondition}
\end{equation}
Moreover, for $\chi=0$, we also find
\begin{eqnarray}
\alpha^{(0)}_{f} = \alpha^{(0)*}_{f*} = 
\alpha^{(0)*}_{b*} = \alpha^{(0)}_{b}\nonumber ,
\end{eqnarray}
which can be used to reduce the numerical effort.

Inspection of the expansion method further reveals that
\begin{equation}
\frac{\partial K_{\chi}}{ \partial \alpha_{i} } =l_{i} ( \boldsymbol \alpha)  
\label{eq:rootEquation}
\end{equation}
for all $i$, which can be used to obtain the  expansion coefficients in Eq.~\eqref{eq:holPrim:firstOrder} using a computer algebra system like \textit{SymPy} or \textit{Mathematica}.

The first total derivative of $\overline K_{\chi} $ with respect to the counting fields gives
\begin{eqnarray}
\frac{d\overline K_{\chi}}{d\chi} =  \frac{\partial \overline K_{\chi}}{ \partial \alpha_{i} } \frac{\partial \alpha_{i} }{\partial\chi}  + 	\frac{\partial \overline K_{\chi}}{\partial \chi},
\label{eq:asympCumGenFkt:firstDer}
\end{eqnarray}
where $\partial/\partial\chi$ and $\partial/\partial \alpha_i$ denote partial derivatives. For brevity, we took advantage of the Einstein notation. Using Eqs.~\eqref{eq:rootEquation} and \eqref{eq:meanFieldCondition}, we find that the first term on the right-hand side vanishes.  Consequently, the first asymptotic cumulant is given by
\begin{eqnarray}
\overline \kappa_1 =  \left.\frac{d\overline K_{\chi}}{d\chi}\right|_{\chi=0} =   \left.	\frac{\partial \overline K_{\chi}}{\partial \chi}\right|_{\chi=0},
\end{eqnarray}
where we recall that the cumulant-generating function has to be evaluated for $\boldsymbol \alpha^{(0)}$.

The second total derivative yields
\begin{eqnarray}
\frac{d^2\overline K_{\chi}}{d\chi^2} =  \frac{\partial \overline K_{\chi}}{ \partial \alpha_{i}\partial \alpha_{j} } \frac{\partial \alpha_{i} }{\partial\chi} \frac{\partial \alpha_{j} }{\partial\chi}  + 2	\frac{\partial \overline K_{\chi}}{\partial \chi\partial \alpha_{j} }\frac{\partial \alpha_{j} }{\partial\chi} +\frac{\partial^2\overline K_{\chi}}{\partial\chi^2}. \nonumber\\
\end{eqnarray}
The evaluation requires the knowledge of $\left.\frac{\partial \alpha_{j} }{\partial\chi}  \right|_{\chi=0}$, which we obtain as follows. Calculating the total derivatives of Eq.~\eqref{eq:rootEquation} with respect to the counting field $\chi$, we obtain
\begin{eqnarray}
0&=& \left. \frac{d\overline K_{\chi}}{d\chi\partial \alpha_{j} }\right|_{\chi=0}  \nonumber \\
&=&  \left.  \frac{\partial \overline K_{\chi}}{  \partial \alpha_{j}\partial \alpha_{i} } \frac{\partial \alpha_{i} }{\partial\chi} \right|_{\chi=0}   + \left. 	\frac{\partial \overline  K_{\chi}}{\partial \chi\partial \alpha_{j} }\right|_{\chi=0} ,
\end{eqnarray}
which is necessarily zero when evaluated for a root $\alpha$. This relation defines an inhomogeneous linear equation, which can be used to obtain the  $\left.\frac{\partial \alpha_{j} }{\partial\chi}  \right|_{\chi=0}$. Eventually,  the asymptotic noise is thus given by
\begin{eqnarray}
\overline \kappa_2  &=& \left. \frac{d^2 \overline K_{\chi}}{d\chi^2} \right|_{\chi=0} \nonumber \\
&=&  \left. 	\frac{\partial \overline K_{\chi}}{\partial \chi\partial \alpha_{j} }\frac{\partial \alpha_{j} }{\partial\chi} \right|_{\chi=0} + \left. \frac{\partial^2 \overline K_{\chi}}{\partial\chi^2} \right|_{\chi=0} .
\end{eqnarray}
Of note, we can obtain the QFI in exactly the same fashion when replacing the derivatives with respect to $\chi$ by derivatives with respect to $\delta$. Likewise, we can also evaluate
\begin{eqnarray}
\frac{d \kappa_1}{d h_z} =\left. 	\frac{\partial \overline K_{\chi}}{\partial \chi\partial \alpha_{j} }\frac{\partial \alpha_{j} }{\partial h_{\text{z}}} \right|_{\chi=0} + \left. \frac{\partial^2 \overline K_{\chi}}{\partial\chi \partial h_{\text{z}}} \right|_{\chi=0},
\end{eqnarray}
where we obtain $\left. \frac{\partial \alpha_{j} }{\partial h_{\text{z}}} \right|_{\chi=0}$ by solving 
\begin{eqnarray}
0&=&   \left.  \frac{\partial \overline K_{\chi}}{  \partial \alpha_{j}\partial \alpha_{i} } \frac{\partial \alpha_{i} }{\partial h_{\text{z}} } \right|_{\chi=0}   + \left. 	\frac{\partial \overline K_{\chi}}{\partial h_{\text{z}} \partial \alpha_{j} }\right|_{\chi=0} ,
\end{eqnarray}
which is a linear inhomogeneious equation.

\begin{figure}
	\includegraphics[width=\linewidth]{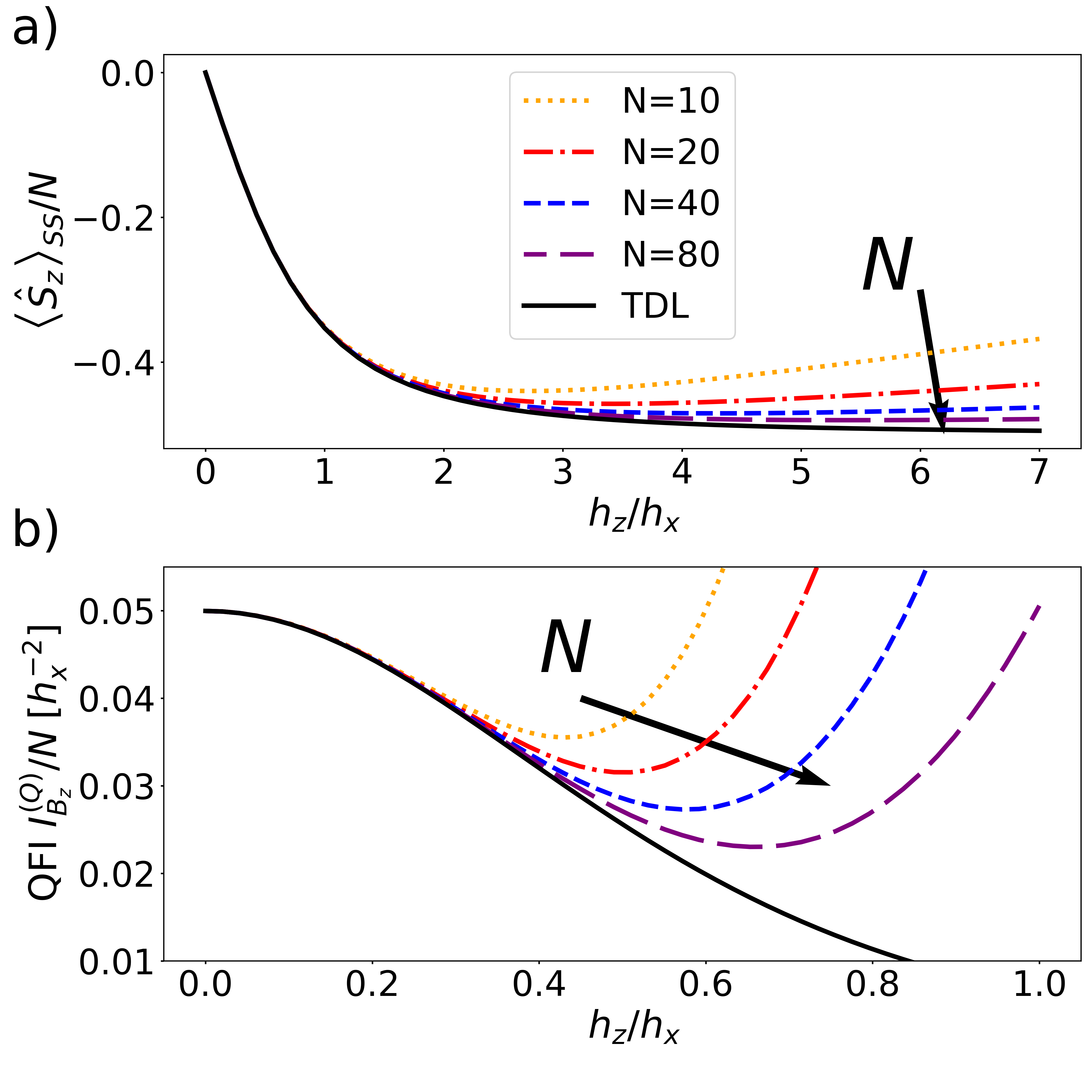}
	\caption{Benchmark calculations of the collective model in Eq.~\eqref{eq:effectiveMasterEquation}. (a) Expectation value of $\hat S_{\text{z}}$ as a function of $h_{\text{z}}$ for four different atom numbers $N=10,20,40,80$ and fixed $\kappa_{\text{z}} = 0$ and $\kappa_{\text{P} }  =0.01 h_{\text{x} } $. The solid black line depicts the mean-field result using the asymptotic cumulant-generating function in Eq.~\eqref{eq:masterEquation:HolsteinPrimakoffMeanfield}. (b) shows the same as (a), but for the QFI.   } 
	\label{figBenchmarkhz}
\end{figure}

\subsection{Benchmark calculations}

\label{app:benchmarking}

\begin{figure*}
	\includegraphics[width=\linewidth]{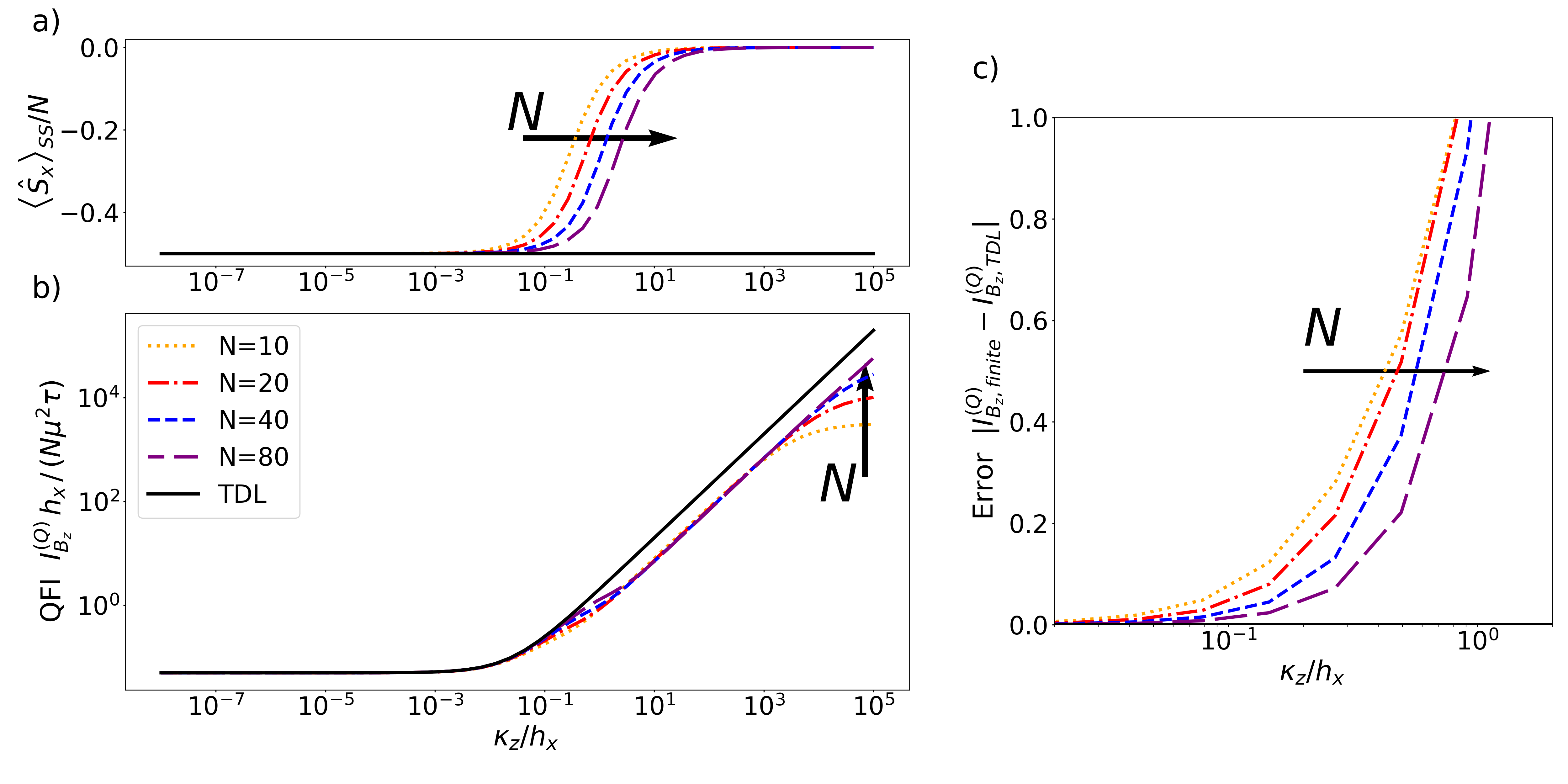}
	\caption{Expectation value of $\hat S_{\text{x}}$ as a function of $\kappa_{\text{z}} $   for $h_{\text{z} } = 0$, $\kappa_{\text{P} }  = 0.1 h_{\text{x} } $  and  atom numbers $N=10,20,40,80$. The solid black line depicts the mean-field result representing the thermodynamic limit (TDL).  (b) QFI  for the finite-size system (colored lines) and in the thermodynamic limit (black). (c) Deviation between the finite-size  and thermodynamic-limit calculations in (b) in the crossover regime.     } 
	\label{figBenchmarkKappa}
\end{figure*}

To validate the non-Hermitian mean-field theory, we compare the  mean-field approach with exact numerical calculations of the finite-size model. To achieve a well-defined  thermodynamic limit, we parameterize the generalized Liouvillian in Eq.~\eqref{eq:collectiveLiouvillian} as follows:
\begin{eqnarray}
\frac{d}{dt} \rho   &=&  -i \left[  \hat {\mathcal H}_{L} \rho-   \rho \hat {\mathcal H}_{R }  \right]  \nonumber \\
&+&  \frac{\kappa_{\text{P} }} {N}  D [  \hat S_{\text{z}} +i   \hat S_{\text{y}}   ] \rho  +	\frac{\kappa_{\text{z}}}{N}   D\left[\hat S_z\right] \rho ,
\label{eq:effectiveMasterEquation}
\end{eqnarray}
For the benchmarking, we will consider the QFI, such that we set $\boldsymbol \chi =0$. Please note that the Hamiltonians $\hat {\mathcal H}_{L}$ , $\hat {\mathcal H}_{R}$ given in Eq.~\eqref{eq:generalizedHamiltonian} still depend on the auxiliary variable $\delta$, required for the evaluation of the QFI. We recall that the QFI can be numerically obtained for finite-size systems by evaluating Eq.~\eqref{eq:FisherInfo:alternative}. For simplicity, we have introduced the effective parameters
\begin{eqnarray}
\kappa_{\text{P} }  &=& N \gamma_{\text{P} }, \nonumber \\
\kappa_{\text{z}}  &=& N \tilde \gamma_{\text{z} } \label{eq:effectiveParameters},
\end{eqnarray}
for which the system in Eq.~\eqref{eq:effectiveMasterEquation} reaches a well-defined thermodynamic limit.

In Fig.~\ref{figBenchmarkhz}(a), we depict the expectation value of $\hat S_{\text{z}}$ as a function of $h_{\text{z}}$ for atom numbers $N=10, 20 , 40, 80$. We observe that the mean-field approach representing the thermodynamic limit $N\rightarrow \infty$ agrees closely to the finite-size calculations for small $h_{\text{z}}$, while there are significant deviations for larger $h_{\text{z}}$. Yet, the finite-size calculation gradually approaches the mean-field result with increasing atom number. In Fig.~\ref{figBenchmarkhz}(b), we investigate the QFI as a function of $h_{\text{z}}$. As in panel (a), the finite-size results converge significantly faster to the mean-field result for smaller $h_{\text{z}}$. Note the different plot ranges in panels (a) and (b), which suggests that fluctuations (such as the QFI) require a higher atom number for convergence than expectation values. Crucially, the mean-field treatment works excellently for $h_{\text{z}}=0$, which is the main focus of this investigation.

In Fig.~\ref{figBenchmarkKappa}, we analyze the convergence of the finite-size calculations as a function of the effective dissipation $\kappa_{\text{z}}$ for $h_{\text{z}} =0$.
In Fig.~\ref{figBenchmarkKappa}(a), we investigate the expectation value of $\hat S_{\text{x}}$. Crucially, the numerical solution of Eq.~\eqref{eq:meanFieldCondition}  is given by  $\boldsymbol \alpha^{(0)} = \boldsymbol 0$, which corresponds to  $\left< \hat S_{\text{x}}\right> /N= -0.5$.   For small $\kappa_{\text{z}}$, the finite-size calculations  have already converged to this value, while for large $\kappa_{\text{z}}$  we observe $\left< \hat S_{\text{x}}\right> = 0 $. For intermediate $\kappa_{\text{z}}$, we find that the crossover point between the $\left< \hat S_{\text{x}}\right> = 0 $ and $\left< \hat S_{\text{x}}\right>/N = -0.5 $ regimes linearly increases towards larger $\kappa_{\text{z}}$ for increasing atom number $N$.

In Fig.~\ref{figBenchmarkKappa}(c),  we depict the QFI as a function of $\kappa_{\text{z}}$ for $h_{\text{z}} =0$. For small $\kappa_{\text{z}}$ we find that the QFI in the thermodynamic limit is independent of  $\kappa_{\text{z}}$ and agrees perfectly with the finite-size calculations.

For intermediate $\kappa_{\text{z}} \approx 0.1 h_{\text{x}}$, we observe finite deviations in the crossover regime from $\left< \hat S_{\text{x}}\right> /N= -0.5$ to $\left< \hat S_{\text{x}}\right> = 0$. As the mean-field approach assumes $\left< \hat S_{\text{x}}\right>/N = -0.5$, the mean-field approach in Sec.~\eqref{sec:nonUnitaryMeanFieldTheory} is invalid, explaining the deviations between the analytical and numerical calculations. Crucially, deviations gradually vanish with increasing atom number. 

For  larger $\kappa_{\text{z}} \approx 0.1 h_{\text{x}}$ we find a linear dependence of both the thermodynamic-limit and finite-size calculations, which differ by a constant factor. For very large $\kappa_{\text{z}} \gg 0.1 h_{\text{x}}$, the finite-size calculations quickly approach to a linear asymptotic behavior. According to  Fig.~\ref{figBenchmarkKappa}(a) and using that $\left< \hat S_{\text{y}}\right> = \left< \hat S_{\text{z}}\right>  = 0$ (not shown), we find that the mean value is located at the center of the Bloch sphere in this exotic asymptotic regime. As  the mean-field approach in App.~\ref{sec:nonUnitaryMeanFieldTheory} can only represent coherent spin states on the surface of the Bloch sphere, it fails to agree with finite-size calculations in this atom-number regime. 

Interestingly, the comparison of the thermodynamic-limit and finite-size calculations in the linear scaling regime for large $\kappa_{\text{z}}$  reveals that both values are identical up to a constant factor $\sqrt{8}$, which we confirmed by comparing various system parameters. Thus,  the large $\kappa_{\text{z }} $ range constitutes a \textit{pre-convergence} regime, in which we find exactly the same scaling properties as in the fully converged system.

\subsection{Integral expression of the photon statistics}

\label{sec:integralExpressions}

In Appendix~\ref{eq:generalizedMasterEquation:collective}, we have developed a formalism to calculate the photonic statistics (i.e., the cumulants) using full-counting statistics. Based on this, we derive here the integral expressions, which are given in the  main text.

Using Eq.~\eqref{eq:lowOrderCumulants}, Eq.~\eqref{eq:momentGenFktExpansion} (which are also valid for the collective model), and Eq.~\eqref{eq:momentGenFkt:collective} we  find  the integral expressions for the first cumulants: 
\begin{eqnarray}
\left< \hat n_\eta \right>  &=&  \int_{0}^{\tau}dt  \left< 	\hat {\mathcal J}_\eta  (t) \right>  +\dot{\overline\nu}_\eta \tau, 
\label{eq:flux:mean}
\end{eqnarray}
where $\dot{\overline\nu}_\eta $  denotes the photon flux of the probe field before the light-matter interaction, and the flux operators $\hat{\mathcal J}_{\eta } = \partial_{\chi_\eta}   \mathcal L_{\boldsymbol \chi=0}  $  are  given by
\begin{eqnarray}
\hat{\mathcal J}_{\text{1} }\rho  &=& \frac{\Omega^2}{8\epsilon_\Delta} \left(\hat S_{\text{z}} \rho + \rho \hat S_{\text{z}}  \right) ,\nonumber \\
\hat {\mathcal J}_{\text{2}}\rho  &=& -\frac{\Omega^2}{8\epsilon_\Delta}  \left( \hat S_{\text{z}} \rho + \rho \hat S_{\text{z}}  \right),
\label{eq:fluxOperators:collective}
\end{eqnarray}
which are superoperators acting on the density matrix.

To obtain the varianace of the photon number difference $\hat n_-$, we derive Eq.~\eqref{eq:momentGenFktExpansion} two times with respect to $\chi_- = \chi_{\text{1}} - \chi_{\text{2}}$ and find
\begin{eqnarray}
\left<  \Delta \hat n_-^2 \right> &=& \int_{0}^{\tau}dt  \left< 	\hat {\mathcal J}^{(2)} (t) \right> \nonumber \\
&+&  \int_{0}^{\tau}dt_1 \int_{0}^{t_1}dt_2  \left<	\Delta \hat {\mathcal J}_-(t_1) 	\Delta \hat {\mathcal J}_- (t_2)\right>  \nonumber \\
&+& \left( \dot{\overline\nu}_{\text{1}} +  \dot{\overline\nu}_{\text{2}}   \right) \tau,
\label{eq:flux:variance}
\end{eqnarray}
where $\Delta \hat {\mathcal J}_-(t)  =  \hat {\mathcal J}_-(t)-\left< \hat {\mathcal J}_-(t) \right> $. Moreover, we have introduced
\begin{eqnarray}
\hat {\mathcal J}^{(2)}  &=& -\partial_\chi^2   \mathcal L_{\boldsymbol \chi=0}, 
\end{eqnarray}
which, however, vanishes for the Liouvillian in Eq.~\eqref{eq:collectiveLiouvillianFinal} due to the adiabatic elimination. Interpreting the superoperators in Eq.~\eqref{eq:fluxOperators:collective} as operators in the Schr\"odinger picture, Eq.~\eqref{eq:flux:mean} and Eq.~\eqref{eq:flux:variance} become Eqs.~\eqref{eq:mean:collective} and \eqref{eq:collectiveNoise}.

\subsection{Exact calculations}

\label{sec:collective:exact}

Here, we evaluate the collective model for $B_{\text{z}} =0$.
Inserting the Bogoliubov transformation of Eq.~\eqref{eq:bogolioubov} into Eq.~\eqref{eq:collectiveMasterEquation} and expanding to the lowest order in the fluctuations we find
\begin{eqnarray}
\frac{d}{dt} \rho   &=&  -i \left[ h_x \hat a^\dagger  \hat a , \rho \right]  \nonumber \\
&+& N  \frac{ \tilde  \gamma_{\text{z}}}{4}  D\left[\hat a + \hat a^\dagger  \right] \rho  + \kappa_{\text{P} }  D [  \hat a  ] \rho 	,
\end{eqnarray}
which is quadratic in the Holstein-Primakoff boson $\hat a$.

In the collective model,  both the measurement noise in Eq.~\eqref{eq:collectiveNoise} and QFI in Eq.~\eqref{sec:quantumFisherInfoAtom} can be expressed in terms of correlation function
\begin{eqnarray}
\int_{0}^\tau dt \left< \Delta \hat S_{z}(t) \Delta \hat S_{z}(0)   \right> + \text{ c.c.} 
\label{eq:fisherInformationLongTimes}
\end{eqnarray}
in the stationary state. As the integrand in Eq.~\eqref{eq:fisherInformationLongTimes} can be expressed using
\begin{eqnarray}
\left< \hat S_{z}(t) \hat S_{z}(0)   \right> &=& \frac{N}{4} \left< \left( \hat a(t) + \hat a^\dagger(t)  \right)   \left( \hat a(0) + \hat a^\dagger(0)  \right)    \right>  \nonumber \\
\end{eqnarray}
and the operator expectation values are given by 
\begin{eqnarray}
\left<\hat a(t) \right>  &=&  e^{\left(  -i h_{\text{x}}  - \frac{ \kappa_{\text{P}}}{2}  \right)t}	\left<\hat a (0 ) \right> \nonumber ,\\
\left<\hat a^\dagger \hat a  \right>_{\text{ss}}  &=&   \frac{ N \tilde \gamma_{\text{z}} }{4 \kappa_{\text{P }} } \nonumber, \\
\left<\hat a  \hat a  \right>_{\text{ss}}  &=&    \frac{  N \tilde \gamma_{\text{z}} }{8 \left(  -i h_{\text{x}}  - \frac{ \kappa_{\text{P}}}{2} \right)  }  ,
\end{eqnarray}
we obtain explicit expressions in  Eqs.~\eqref{eq:quantumNoiseCollective}  and ~\eqref{eq:quantumFisherInfoCollective} using the  quantum regression theorem.

{ \color{\markColorTwo}

\subsection{Experimental limitations}

\label{sec:experimentalLimitations}

 In Sec.~\ref{sec:collectiveModel}, we have  investigated the fundamental properties of the quantum information in the collective model. Here, we add a discussion about a series of other experimental circumstances, which might obscure the observation of the Heisenberg scaling.
\begin{itemize}
	\item \textit{Absorption and spontaneous photon emission.} For a very  large atom number, the absorption of the probe field will become significant. This will be accompanied with an increased spontaneous photon emission, which can be obstructive to the  collective measurement-induced correlations. Yet, this effect can be mitigated by choosing a large detuning $\epsilon_\Delta$ accompanied by a stronger probe field, which in effect leads to a smaller absorption, while the effective dissipation $ \tilde \gamma =  \tilde \gamma_{z}  =  \frac{\Gamma\Omega^2}{\epsilon_\Delta^2 +  \frac{\Gamma^2}{4} }  $ in Eq.~\eqref{eq:collectiveMasterEquation} remains unchanged.
	
	\item \textit{Atomic motion.} As the atomic motion is independent of the magnetic state of the atoms, its effect is vanishing small. Due to the large probe laser detuning, the polarization rotation is independent of the detuning, which renders the protocol insensitive to the Doppler effect.
	
	\item \textit{Technical noise.} Realistic experiments might suffer from technical noise, such as laser noise or magnetic field noise. Yet, as these noise sources are not fundamental, they might be reduced by engineering solutions to such a level that the Heisenberg scaling can be isolated from the measurement data.
	
	\item \textit{Atomic interactions.} For high atom densities, atomic interactions will become relevant. In how far this will benefit or obstruct the QFI and the SNR must be assessed by deploying more sophisticated models in future research. Similarly, models which microscopically describes the non-collective action of the pump term will be necessary to investigate the resilience of the predicted Heisenberg scaling.
	
\end{itemize}
}

\bibliography{projectLibrary}

\end{document}